\documentclass[numberedappendix,appendixfloats,iop,apj,revtex4]{emulateapj}

\usepackage[usenames,dvipsnames]{color}
\usepackage[colorlinks=true, anchorcolor=blue, linkcolor=blue, urlcolor=Fuchsia, citecolor=NavyBlue, breaklinks=true]{hyperref}
\usepackage{times, apjfonts}
\usepackage{amsmath, amsthm, amssymb}
\usepackage{graphicx, epsfig}
\usepackage{natbib}
\usepackage{epstopdf} 
\usepackage{CJK} 
\usepackage[hyphenbreaks]{breakurl}

\shorttitle{Broad H$\beta$ Emission-Line Variability}
\shortauthors{Runco et al.}

\begin{document}

\title{Broad H$\beta$ Emission-Line Variability in a Sample of 102 Local Active Galaxies}

\author{Jordan N. Runco\altaffilmark{1}, Maren Cosens\altaffilmark{1}, Vardha N. Bennert\altaffilmark{1}, Bryan Scott\altaffilmark{1}, S. Komossa\altaffilmark{2},
Matthew A. Malkan\altaffilmark{3},
Mariana S. Lazarova\altaffilmark{4},
  Matthew W. Auger\altaffilmark{5}, Tommaso Treu\altaffilmark{3}, \and Daeseong Park\altaffilmark{6}}
\affil{$^{1}$Physics Department, California Polytechnic State University, San Luis Obispo
   CA 93407, USA; jrunco@calpoly.edu, mcosens@calpoly.edu, vbennert@calpoly.edu}
\affil{$^{2}$Max-Planck-Institut f\"ur Radioastronomie, Auf dem H\"ugel 69, 53121, Bonn, Germany}
\affil{$^{3}$Department of Physics, University of California, Los Angeles, CA 90095, USA; malkan@astro.ucla.edu, tt@physics.ucsb.edu}
\affil{$^{4}$Department of Physics and Physical Science, University of Nebraska Kearney, Kearney, NE 68849; lazarovam2@unk.edu}
\affil{$^{5}$Institute of Astronomy, Madingley Road, Cambridge CB3 0HA, UK; mauger@ast.cam.ac.uk}
\affil{$^{6}$Korea Astronomy and Space Science Institute, Daejeon, 34055, Republic of Korea; daeseongpark@kasi.re.kr}

  \begin{abstract} 
    A sample of 102 local ($0.02 \leq z \leq 0.1$) Seyfert galaxies with black hole masses $M_{\rm{BH}} > 10^{7}M_{\odot}$ was selected from the Sloan Digital Sky Survey (SDSS) and observed using the Keck 10-m telescope to study the scaling relations between $M_{\rm{BH}}$ and host galaxy properties.  We study profile changes of the broad H$\beta$ emission line within the $\sim3-9$ year time-frame between the two sets of spectra.  The variability of the broad H$\beta$ emission line is of particular interest, not only since it is used to estimate $M_{\rm{BH}}$, but also since its strength and width is used to classify Seyfert galaxies into different types.
    At least some form of broad-line variability (in either width or flux)
    is observed in the majority ($\sim$66\%) of the objects, 
    resulting in a Seyfert-type change for $\sim$38\% of the objects,
    likely driven by variable accretion and/or obscuration. 
    The broad H$\beta$ line virtually disappears in 3/102 ($\sim$3\%) extreme cases.
    We discuss potential causes for these changing-look AGNs.  While similar dramatic transitions have previously been reported in the literature, either on a case-by-case basis or in larger samples
    focusing on quasars at higher redshifts, our study provides statistical information  on the frequency of H$\beta$ line variability in a sample of low-redshift Seyfert galaxies.
    \newline
\textit{Subject headings:} accretion, accretion disks $-$ black hole physics $-$ galaxies: active $-$ galaxies: evolution $-$ galaxies: Seyferts $-$ galaxies: statistics
\end{abstract}

\section{Introduction}
\label{intro}
Observed relations between the mass of the supermassive black hole
($M_{\rm{BH}}$)
at the center of a galaxy
and the properties of its host galaxy --
such as host galaxy mass \citep[][]{mag98}, luminosity \citep[]{kor95}, and stellar velocity dispersion \citep[]{fer00, geb00} --
imply a relationship
between galaxy evolution and black hole (BH) growth
\citep[for a recent review see][and references therein]{kor13,gra16}.
In the local Universe, $M_{\rm{BH}}$ can be measured by spatially resolving the BH sphere of influence using stellar or gas kinematics \citep[e.g.,][]{van98,geb00}.  At larger distances, 
the only way to estimate $M_{\rm{BH}}$ is 
by resolving the BH sphere of influence as it responds to variations in continuum in galaxies with an active galactic nucleus (AGN).
In AGNs, the BH is actively growing via
an accretion disk of in-falling material.
The high energy photons emitted by the hot accretion disk ionize the surrounding
gas clouds, the broad-line region (BLR) in the vicinity of the BH, and the
narrow-line region (NLR) further out.
Reverberation mapping \citep[]{wan99, kas05, ben13}
traces variations in the accretion-disk continuum luminosity 
and the time-delayed response of the BLR flux to determine
the size of the BLR, using light-travel time arguments.
The velocity of the BLR gas can be determined from the Doppler broadening
of the emission lines (such as the broad H$\beta$ line in the rest-frame optical).
By assuming a dimensionless virial coefficient to describe the kinematics and geometry of the BLR, the velocity and size of the BLR combined yield $M_{\rm{BH}}$.
More recently, there have been attempts to estimate $M_{\rm{BH}}$ in individual objects independent of a virial coefficient
by modeling reverberation-mapped data directly and
constraining the geometry and kinematics of the BLR
\citep[see e.g.,][and references therein]{pan14}. 

Seyfert galaxies are low luminosity AGNs for which the host galaxy can be easily resolved,
thus making them attractive targets for the study of the $M_{\rm{BH}}$ scaling relations.
In fact, for a few Seyfert galaxies, $M_{\rm{BH}}$ estimates from both dynamical AO measurements
as well as reverberation mapping are available \citep[][]{hic08}.
Seyfert galaxies are categorized into different types based on their emission line profiles, ranging from type-1 to type-2 with subclasses (type-1.5, 1.8, and 1.9) in-between.  Type-1 Seyferts display both broad and narrow components of emission lines, while type-2 Seyferts show only the narrow components.  The intermediate Seyfert types show varying levels of broad component emission.  
The Balmer series in the optical regime is generally used to classify
Seyfert type \citep[][Table~\ref{classification} summarizes the different Seyfert-type classifications]{ost76, ost77, ost81}.

 In the framework of the so-called standard unified model for active galaxies, all Seyfert galaxies are thought to be intrinsically the same but viewed from a different angle.
 The key to this model is a region of cold gas and dust, called the dusty torus,
 that surrounds the BLR and, if seen edge on, can shield both the accretion disk continuum
 and the broad emission lines from the observer's view, resulting in a type-2 Seyfert galaxy.
If seen face on, however, both accretion disk and BLR are visible, resulting in a
 type-1 Seyfert galaxy.
 Intermediate types-1.5, 1.8, and 1.9 are viewed along the edges of the dusty torus where it is not optically thick enough to fully block the broad lines.
 In other words, in this simplified model, the presence or absence of broad lines is attributed
 solely to viewing orientation, meaning that the Seyfert type of a galaxy does not change.
 (Note that the orientation of the torus is entirely independent of the host-galaxy orientation.)

 However, there have been many reports of apparent Seyfert-type changes in the literature
 \citep[e.g.,][]{toh76,kol85,sto93,are99,era01,tri08,den14,sha14,park15}.
These changes can occur in either direction.
NGC 4151 is one of the most notable and cited examples.  Originally classified as a type-1.5 \citep[][]{ost77}, the broad emission lines disappeared throughout the 1980's \citep[][]{ant83, lyu84, pen84}, but have returned since \citep[][]{sha10}.  Another well-studied example is Mrk 590 which has been observed over a 40-year time scale.  First observed as a type-1.5, Mrk 590 transitioned to a type-1 before the broad lines disappeared, making it a type$\sim$1.9-2 Seyfert \citep[][]{den14}.
Possible causes of these changing-look AGNs include changes in extinction
\citep[due to our line-of-sight grazing the dusty torus, for instance; e.g.,][]
      {goo89,lei15}, or changes in the AGN accretion rate
\citep[e.g.,][]{nic00,kor04,eli14}
In rare cases, the increase in accretion rate could be due to the tidal
disruption and accretion of a star, and a few cases of dramatic broad-line
variability possibly linked to this scenario, have been reported in recent years
\citep[e.g.,][]{kom08,arc14,lam15,mer15}.
 
A certain degree of variability in the flux and profile of the BLR emission,
as a response to changes in the continuum flux (and thus accretion),
is not only expected, but in fact forms the basis for reverberation mapping studies.
Such studies have shown that variations in continuum flux and that of the broad Balmer lines
are correlated in a way so that the derived $M_{\rm{BH}}$ does not change \citep[][]{ben07, par12, bar15}. 
However, extreme variability leading to a type change seems to be rare.

 Here, we address the question of the frequency of these changing-look AGNs
 by taking advantage of a statistical sample of 102 local Seyfert galaxies with
 archival spectra from the Sloan Digital Sky Survey (SDSS) and high-quality
 Keck spectra taken $6.4 \pm 1.8$ years apart.
 The paper is organized in the following manner.  Section~\ref{sample} summarizes the sample selection, observations, and data reduction.  Section~\ref{analysis} describes the analysis
 of the data. Section~\ref{results} discusses
 the derived quantities and results from the data.
 Section~\ref{summary} concludes with a summary.  The appendix contains a table of observations and Seyfert-type transitions for the sample (\ref{appendix:observations}), comparison of SDSS and Keck spectra (\ref{appendix:agnspectra}),
 fits to the SDSS and Keck spectra (\ref{appendix:fits}),
 and SDSS optical images
 for two extreme objects (\ref{appendix:sdss}).
Throughout the paper, a Hubble constant of H$_{\rm{o}}$ = 70 km s$^{-1}$, $\Omega_{\rm{\lambda}}$ = 0.7, and $\Omega_{\rm{M}}$ = 0.3 is assumed.

\section{Sample Selection, Observations, and Data Reduction}
\label{sample}
The primary goal behind sample selection and observations
is the creation of a local baseline for the BH mass scaling relations
of active galaxies, presented by \citet{ben11,har12,ben15}.
These papers describe the sample selection, Keck observations, and Keck data reduction in detail. We here provide a brief summary only.

\subsection{Sample Selection}
A sample of 102 local (0.02 $\leq z \leq$ 0.1) type-1 Seyfert galaxies  was selected from the SDSS data release six (DR6) \citep[][]{ade08}.  Objects were selected on the basis of a broad H$\beta$ emission line with an estimated $M_{\rm{BH}}$ \textgreater $10^{7}M_{\odot}$ \citep[][]{ben11, har12}.  Note that of these 102 objects, only 79 are used by
\citet{ben15} to study the M$_{\rm{BH}}$-$\sigma$ relation,
since the necessary quantities (i.e., $\lambda$L$_{\rm{5100}}$, M$_{\rm{BH}}$, $\sigma$)
for that study were only accessible for these 79 objects.

\subsection{SDSS Observations and Data Reduction}
SDSS spectra are obtained from a 2.5-m ground based telescope with a 3" diameter circular optical fiber and an exposure time of 54 seconds.  SDSS spectra cover a wavelength range of 3800\AA \space to 9200\AA \space with an instrumental resolution of 170 km s$^{-1}$.
SDSS data are already fully reduced and flux calibrated when retrieved from the SDSS archive. 

\subsection{Keck Observations and Data Reduction}
\label{keck}
The 102 objects selected from SDSS were observed again between January 2009 and March 2010 with the Low Resolution Imaging Spectrometer (LRIS) at the Keck 10-m telescope using a 1" x 2" wide rectangular long slit aligned with the major axis of the host galaxy (given by SDSS).
(While all objects were observed at as low airmass as possible,
  given observation constraints, 
  for individual objects the airmass can be as high as 1.4.) Objects observed in 2009 used a D560 dichroic, and objects observed in 2010 used a D680 dichroic.  The blue Keck spectra were taken with the 600/400 grism giving a wavelength range of $\sim$3200-5350\AA \space and an instrumental resolution of $\sim$90 km s$^{-1}$; the red spectra were taken with the 831/8200 grating centered on 8950\AA \space with a resolution of $\sim$45 km s$^{-1}$.  (Note that the red Keck spectra are not used in this paper since they only cover the Ca triplet absorption lines for an accurate measurement of $\sigma$).  The exposure times generally range from 600 to 1200 s. 
Keck spectra were taken on average
6.4 $\pm$ 1.8 years after the SDSS spectra,
ranging from 2.6 to 9.1 years (see Table~\ref{table:observations} for
details on SDSS and Keck observations).

The Keck data are reduced following standard reduction steps such as bias subtraction, flat field correcting, cosmic ray rejection, and wavelength calibration. AOV Hipparcos stars were used to correct for telluric absorption and relative flux calibration.  Note that, unlike SDSS spectra, Keck spectra are not absolute flux calibrated because observing conditions were typically not photometric.
1D spectra were extracted from the 2D spectra with a width of 1.08" (8 pixels) to encompass the BLR,
given the slit width of 1'' and a typical seeing of 1''.

\subsection{Lick Observations and Data Reduction}
For eight objects with significantly weaker or apparently absent broad H$\beta$ emission in the Keck spectra,
follow-up observations were conducted in January and March of 2013
with the 3m Shane telescope of Lick observatory
using the Kast spectrograph and
60 min total exposure time per object.
\citep[Table~\ref{table:observations};][]{sco13}.  The slit was aligned either along the major axis or perpendicular to it.  1D spectra were extracted using a 4 pixel ($\approx$3") width centered on the peak flux to mimic the 3" diameter circular fiber of SDSS.  The data were reduced following standard procedures.
The Lick spectra are presented by \citet{sco13}.
These spectra are used to determine their Seyfert type,
since they also cover the H$\alpha$ region. The Seyfert types based on the Lick spectra are listed
in Table~\ref{table:observations}.

\section{Analysis}
\label{analysis}
In this paper, we focus on four different sets of spectra,
for short called ``Keck subtracted,'' ``Keck unsubtracted,'' ``SDSS subtracted,'' and ``SDSS unsubtracted,'' as explained in the next two sections.

\subsection{Unsubtracted Spectra}
\label{unsubtracted}
To classify Seyfert type and perform a qualitative comparison of the H$\beta$ region,
the reduced spectra are used.
Throughout the paper, these spectra are referred to as the unsubtracted data set.

For a visual comparison of both data sets (Appendix~\ref{appendix:agnspectra}),
the Keck spectra were re-binned to match the lower spectral
resolution of the SDSS spectra.
For visual presentation purposes of a direct comparison between both spectra
in Figures~\ref{appendix:agnspectra},
the spectra were normalized to constant 5007\AA \space [OIII] emission,
assuming that the 5007\AA \space [OIII] emission line flux is identical in both data sets,
given that emission from the extended NLR does not vary over the observed timescales. 
This scaling also assumes that both spectra integrate the same [OIII] emission
over the same area which might not necessarily be the case, given the different apertures used.
Aligning the long Keck slit along the major axis of the host galaxy may reduce
any difference in [OIII] flux covered.
However, we discuss aperture effects in more detail below.

\subsection{H$\beta$ Fitting and Subtracted Spectra}
\label{subtracted}
A multi-component spectral decomposition is used to fit the region around H$\beta$.
The procedure is summarized here briefly \citep[see][for details]{par15}.

First, the observed continuum is modeled and subtracted by fitting a pseudo-continuum consisting of the featureless AGN power-law continuum, host-galaxy starlight templates from the Indo-US spectral library \citep[][]{val04}, and the AGN FeII emission template from \citet{bor92} for the Keck spectra and from \citet{kov10} for the SDSS spectra\footnote{The two different FeII templates are being used due to the different wavelength ranges covered by the Keck and SDSS spectra. The Keck spectra do not extend far enough into the red to fit the FeII features around 5200\AA~with the \citet{kov10} multi-component template and instead must be fitted with the monolithic template from \citet{bor92}.}.  Then, the continuum-subtracted H$\beta$ line region is modeled by fitting Gauss-Hermite series \citep[][]{mar93, woo06, mcg08} simultaneously to the [OIII] narrow emission lines $\lambda\lambda$4959,5007\AA~and the broad and narrow H$\beta$ lines, to allow for the fitting of
asymmetries. Gauss-Hermite polynomials of order 3-6 are used to fit the broad H$\beta$ line and 7-12 for each [OIII] line. In cases where the HeII $\lambda$4686\AA~emission line is blended
with the broad H$\beta$, the broad and narrow HeII were fitted by simple Gaussian functions.

Depending on the degree of overlap of the broad H$\beta$ component with the [OIII] lines, we model the H$\beta$ line region in two slightly different ways.
If there is no blending between the broad H$\beta$ component and the [OIII]$\lambda$5007\AA~line,
we create a template for the narrow line components by fitting the [OIII]$\lambda$5007\AA~line
with a Gauss-Hermite series function. The [OIII]$\lambda$4959\AA~line is then subtracted by blueshifting the template with a flux scale ratio fixed to 1:3 \citep[see, e.g.,][and references therein]{dim07}.
Then, the broad and narrow H$\beta$ components are fitted simultaneously through $\chi^2$-minimization,
using the blueshifted template from the [OIII]$\lambda$5007\AA~line as a template for the narrow H$\beta$,
with the flux ratio as a free parameter, and a Gauss-Hermite series for the broad H$\beta$ component.

If the broad H$\beta$ component is heavily blended with the [OIII] doublet lines, we model the H$\beta$ broad and narrow lines and the [OIII]$\lambda\lambda$4959, 5007\AA~lines all together by simultaneously fitting a Gauss-Hermite series function to the broad H$\beta$ component and another Gauss-Hermite series function to the [OIII]$\lambda$5007\AA~line, where the model for the [OIII]$\lambda$5007\AA~line is blueshifted and also used for both the [OIII]$\lambda$4959\AA~line with a 1:3 flux scale ratio and the narrow H$\beta$ component with a free flux scale ratio. This approach is based on the known fixed flux ratio of [OIII]$\lambda\lambda$4959, 5007 of 1:3
and the fact that the [OIII] lines and the narrow H$\beta$ lines originate in the NLR and should have
comparable widths. It is an approach typically used to fit AGN spectra in e.g., reverberation mapping studies
\citep[][]{par15,bar15}.

The results of the spectral fitting are given in Table~\ref{table:fitresults}, fits are shown in Appendix~\ref{appendix:fits}.
The pure emission-line spectra -- with host-galaxy, power-law continuum and FeII emission subtracted --
are referred to in the following as the subtracted data.

Note that for one object  in the sample (1655+2014), the S/N ratio is too low (in both SDSS and Keck spectra)
for an accurate measurement of the broad H$\beta$ line.
For three other objects (0932+0405, 0847+1824, and 0831+0521),
the broad H$\beta$ component could not be accurately identified in the Keck spectra.
These four objects are excluded from any discussion involving the H$\beta$ line fitting.

\subsection{Seyfert-Type Classification}
The Seyfert type for all objects was classified independently by eye by two members of the team
(JNR \& MC) following Table~\ref{classification},  and has been verified independently
by the broad H$\beta$ emission-line fitting results.
Table~\ref{table:observations} lists the Seyfert-type classifications for both SDSS and Keck spectra, as well as the eight objects observed at Lick.
Typically, the H$\alpha$ and H$\beta$ lines are used for Seyfert-type classification,
and we followed that procedure for the SDSS and Lick spectra.
However, the Keck spectra do not extend to the H$\alpha$ line.
Instead,
higher order Balmer lines (H$\gamma$ and H$\delta$)
were used as a proxy, see Table~\ref{classification} \citep[][]{ost77}.
However, these lines are intrinsically much fainter than H$\alpha$ and H$\beta$;
for example, assuming case B recombination,
H$\delta$ (H$\gamma$) is $\sim$26\% (47\%) the strength of H$\beta$
which itself is approximately 35\% of H$\alpha$ \citep[][]{ost89}.
Moreover, the H$\gamma$ line is often blended with the 4383\AA \space FeI and the 4363\AA \space [OIII] lines, cautioning the use of this line for classification.
Thus, we rely on the  H$\beta$ line for classification.
This implies that we cannot differentiate between types-1.9 and 2 for Keck spectra.
We conservatively classify an object without a broad H$\beta$ line in the Keck spectra
as a type-1.9.
Note that given the lower S/N of the SDSS spectra, caution should be exercised when classifying the Seyfert sub-types (1.5, 1.8, and 1.9) since broad lines can be easily lost in the noise. However,
given that these are all local Seyfert galaxies, generally, S/N ratios are good even for the SDSS spectra.
The observed variability of the broad H$\beta$ line is substantial and the
overall trend is for a weaker broad H$\beta$ line in Keck, partially due to selection effects (see discussion below).

For a more quantitative determination of Seyfert type, we used the H$\beta$ broad/narrow peak-flux ratios
and compared them to the visual classification.
The reason for choosing peak-flux ratios
are threefold. (i) Using flux ratios eliminates uncertainties on absolute flux calibration. Moreover, compared to integrated flux ratios,
(ii) peak-flux ratios are essentially driving the visual classification scheme; and (iii) the width of the broad H$\beta$ line can have large uncertainties and depends strongly on the placement of the continuum. Depending
on the signal-to-noise (S/N) of the data, it can easily be either lost in the noise or
noise can be fitted as a broad H$\beta$ line. Thus, integrated-flux ratios are more uncertain than peak-flux ratios.

While there is scatter between these two different
classification approaches, we determined cutoffs in the H$\beta$ broad/narrow peak flux ratios by
minimizing the number of outliers. For a peak-flux ratio $\geq$ 1.25 there is an 83\% chance that the object is a type-1, for 1.25 $\geq$ peak-flux ratio $\geq$ 0.6, there is a 67\% chance that the object is a type-1.5, and for a peak-flux ratio $\leq$ 0.6 there is a 72\% change that the object is a type-1.8.
All type-1.9 objects have a peak-flux ratio of 0, since there is no broad H$\beta$ component.
Below (Section~\ref{transitions}), we discuss the results based on both visual classification as well as peak-flux ratios.
The peak-flux ratios are given in Table~\ref{table:observations}.

\begin{deluxetable*}{ll}
\tabletypesize{\scriptsize}
\tablecolumns{2}
\tablecaption{Seyfert-Type Classifications}
\tablehead{
  \colhead{Type} &
  \colhead{Description}\\
  \colhead{(1)} & \colhead{(2)}}
\startdata
  Type-1 & Both broad and narrow components in all Balmer lines. \\
 Type-1.5 & Broad and narrow components can be identified in H$\alpha$ and H$\beta$.  Broad
component of higher order Balmer lines are weakening. \\
 Type-1.8 & Broad H$\beta$ is weak but detectable.  No higher order Balmer lines have a broad component. \\
 Type-1.9 & Shows broad H$\alpha$ but no higher order Balmer lines have a broad component. \\
 Type-2 & No broad emission lines. 
 \enddata
 \tablecomments{
   Col. (1): Seyfert type.
   Col. (2):
   Seyfert-type classification based on the strength of the H$\beta$ and H$\alpha$ lines \citep[][]{ost77, ost81}.}
 \label{classification}
\end{deluxetable*}

\section{Results and Discussion}
\label{results}

\subsection{Seyfert-Type Transitions}
\label{transitions}
\begin{figure}[ht!]
\centering
\includegraphics[width=8cm, height=8cm]{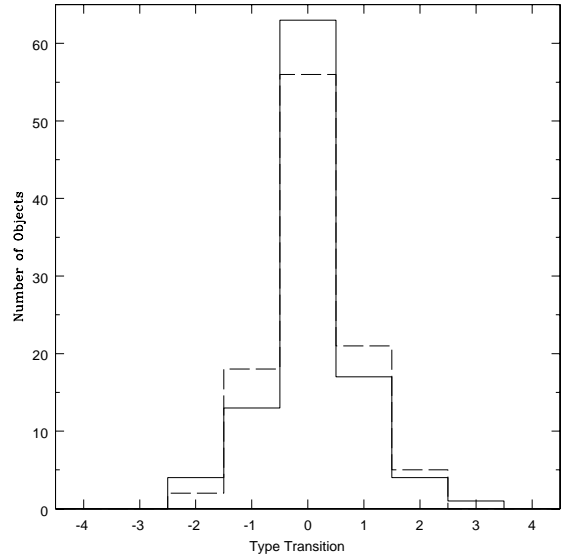}
\caption[Magnitude of Seyfert-type transitions.]{Magnitude of Seyfert-type transitions
from visual classification (solid line) and peak-flux ratio (dashed line; see text for details).}
\label{fig:histogram}
\end{figure}

Based on our visual classifications,
at least some degree of type transition is exhibited by
39/102 objects (39$\pm$10\%).
To quantify the ``magnitude'' of the transition,
we assign a value between +4 and $-$4 in increments of 1,
with a positive value if the broad H$\beta$ line weakened
between SDSS and Keck. Type changes of
+4 indicate a full type transition from type-1 to 2; +3 indicates the object transitioned three types (e.g., 1 to 1.9); +2 indicates a transition of two types (e.g., 1 to 1.8); and +1 indicates a transition of one type (e.g., 1 to 1.5); 0 indicates that the object did not experience a Seyfert-type change.
A negative value implies that the H$\beta$ line increased between SDSS and Keck.
Figure~\ref{fig:histogram} shows the distribution of the type transitions quantified in this way.

While there are type transitions in either direction, 
there are more objects that transition towards a narrower/weaker broad H$\beta$ line in Keck (transition towards type-2) which is likely a reflection of our sample selection,
since only Seyfert galaxies with a broad H$\beta$ component were selected from SDSS \citep[][]{ben11, har12}.
Of the 39 objects with a type transition, the majority (31) experiences only
a minor type change ($\pm$1 magnitude transition), while eight underwent greater type changes of 2.
One object (0847+1824) demonstrated a type transition of magnitude 3.
We consider three objects (3\% of the sample) extreme objects
since they all show cases of a disappearing
broad H$\beta$ line between SDSS and Keck. We discuss them in detail in Section~\ref{extreme}.
There is no correlation between the magnitude of the type transition and the time between observations, indicating that a Seyfert-type change happens on shorter time scales than those covered by our observations (6.4$\pm$1.8 years).  

If we instead use our peak-flux ratio cutoffs, as discussed in Section~\ref{transitions}, to classify the Seyfert type, the results
change only slightly. 46/102 objects undergo a type transition, with the majority (39) experiencing only
a minor type change ($\pm$1 magnitude transition), while 7 underwent greater type changes of 2.
However, we consider the visual classification more reliable, and it is also commonly used in the literature.
Thus, in the following we refer to the visual classification.

As outlined in the introduction,
AGN type changes have previously been observed in the
literature and are often referred to as "changing-look" AGNs
\citep[see, e.g.,][and references therein]{den14,lam15,run16};
indeed, the frequency and strength of our observed H$\beta$ emission line variability
is in line with one of the first studies on this topic
\citep[][]{ros94}.

Two common explanations of type transitions are variable accretion and variable obscuration.  Variable accretion, caused by fluctuating amounts of gas available to feed the BH \citep[][]{boc06}, results in a change of the continuum flux and subsequently in
a change of the broad H$\beta$ emission line flux, since the BLR clouds are photoionized by the UV continuum.
There are many papers citing variable accretion as the driving force behind Seyfert-type transitions
\citep[][]{era01,tri08,sha14,den14}. \citet{den14} is the most notable example reporting 
Mrk 590 to transition from type-1.5 to type-1, then transition again to type$\sim$1.9-2.  \citet{sha14} (for NGC\,2617), \citet{tri08} (for NGC\,2992), and \citet{ant83} (for NGC\,4151) report an observed change in X-ray flux that is followed by a similar change in UV/optical flux.
Variable extinction can occur when dusty clouds pass our line-of-sight. 
For Seyfert galaxies, the source of this obscuration is likely the dusty torus, thought to surround the BLR in the framework of the standard unified model.  The individual cold gas clouds of the dusty torus are not all identical and uniform, so different gas clouds could shield different amounts of continuum and H$\beta$ flux.

The disappearance of broad lines as seen in 3\% of our sample
(discussed in detail in Section~\ref{extreme}) has
also been documented before. In particular,
\citet{ho09} report that ten of the 94 objects in their local sample ($\sim$ 11\%),
selected to investigate relationships between $M_{\rm{BH}}$~and host galaxy properties
using Magellan spectra (3600-6000\AA~wavelength range),
had only narrow lines when the same objects were previously classified as type-1 Seyferts.
(Note that \citealt{ho09} do not discuss
this matter further, since it was not the main purpose of their paper.)

In a sample of 117 changing-look quasar candidates selected from SDSS DR12,
\citet{rua15} discovers two new low-redshift quasars (in addition to the one previously
found by \citealt{lam15})
where both the broad H$\beta$ and continuum luminosity dim over the 5$-$7 years in rest-frame time, changing the
objects from quasar-like to galaxy-like states.
\citet{rua15} argue that the observed change is driven by rapidly decreasing accretion rates.

From a sample of 1011 low-redshift quasars (z $<$ 0.63), selected based on
repeated photometry from SDSS and Pan-STARRS1 as well as repeated spectra from SDSS and SDSS-III Boss,
\citet{mac15} visually identify ten changing-look quasars, with 4/10  of these objects showing emission lines broadening with at least a one magnitude increase in g-band flux and  5/10 of these objects having disappearing broad emission lines and a decreasing light curve (one of these five objects was discovered by \citealt{lam15}).
One of these five objects (SDSS J1021+4645) experienced a type transition from type-1 to type-1.9.
\citet{mac15} report significant changes on timescales of $\sim$2000-3000 days with broad emission line changes corresponding to continuum changes.  Variable accretion and obscuration were both discussed as possible options to explain the observed broad-line changes, and neither possibility was ruled out \citep{mac15}.
A tidal disruption flare event might explain the observed changes behind J0159+0033
\citep{mac15, mer15}.  

For the purpose of a reverberation mapping campaign,
\citet{bar15} re-observed AGNs classified as Seyfert-1 galaxies
based on SDSS spectra 5-8 years later with the 
3m Shane telescope of Lick observatory and noted that for one object
(NGC 6423), all emission lines had disappeared, and three other objects (Mrk 474, Mrk 728, and Mrk 1494) changed from Seyfert-1 to Seyfert 1.9.
We will discuss possible explanations for such extreme changes in Section~\ref{extreme}.

\subsection{Quantifying the Observed Broad H$\beta$-Line Changes}
\label{quantify}
\begin{figure}[h!]
\centering
\includegraphics[width=\columnwidth]{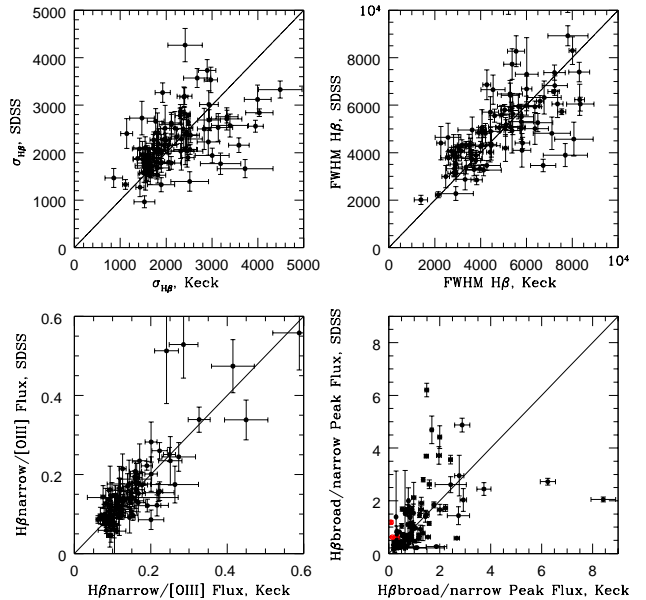}
\caption{Derived quantities from H$\beta$ line fitting of SDSS spectra
  (y-axis) vs. Keck spectra (x-axis; including a unity line).
  Top left: $\sigma_{\rm{H\beta}}$.  Top right: FWHM$_{\rm{H\beta}}$.  Bottom left: H$\beta_{\rm{narrow}}$/[OIII] flux ratio.  Bottom right: H$\beta_{\rm{broad}}$/H$\beta_{\rm{narrow}}$ peak flux ratio.
  The two objects shown in red in this panel were classified as 1.9 in the Keck spectrum, thus they do not have a broad component in H$\beta$ and this comparison uses an upper limit on the flux if a broad component were included.
}
\label{fig:ratios}
\end{figure}

We use several measurements
from our spectral decomposition of the H$\beta$ region discussed in Section~\ref{subtracted}
to further quantify the observed broad H$\beta$ line changes and explore possible
origins,
particularly in these four quantities:  (i) the second moment of the broad H$\beta$ component ($\sigma_{\rm{H\beta}}$) from the model, used to calculate $M_{\rm{BH}}$ \citep[][]{ben11, ben15}
(ii) the full width at half maximum (FWHM) of the broad H$\beta$ line also sometimes
used to calculate $M_{\rm{BH}}$ \citep[e.g.,][]{she08,she11}\footnote{For a Gaussian profile, FWHM$_{H\beta}$ = 2.35$\sigma_{\rm{H\beta}}$.}
(iii) the flux ratio of the narrow H$\beta$ component and [OIII] lines, H$\beta_{\rm{narrow}}$/[OIII]
(iv) the peak-flux ratio of the broad and narrow components of the H$\beta$ line, H$\beta_{\rm{broad}}$/H$\beta_{\rm{narrow}}$.
In Figure~\ref{fig:ratios}, we compare these four quantities
as derived from the SDSS spectra  with those derived from the Keck spectra.

On average, the broad H$\beta$ line is wider in the SDSS spectra than in the Keck spectra, both in $\sigma_{\rm{H\beta}}$ (1.07$\pm$0.29) and FWHM$_{\rm{H\beta}}$ (1.08$\pm$0.27) (Table~\ref{statistics}). Moreover, the broad H$\beta$ line has more peak flux in the SDSS spectra
than in the Keck spectra (1.26 $\pm$ 0.12).  
The broader  and stronger H$\beta$ line preferentially in the
SDSS spectra is likely attributed to the bias in the sample selection,
since only objects with a broad H$\beta$ line were chosen from SDSS.

What is noticeable in the top left panel of Figure~\ref{fig:ratios}
is the large scatter: for individual objects,
$\sigma_{\rm{H\beta}}$ can be almost
up to a factor of two different between the two sets of spectra. FHWM follows a similar trend.
Reverberation mapping studies have shown that
the variability of the line-width in AGNs
correlates inversely with the variability of the power-law continuum
in a way that it cancels out in the virial product,
resulting in a constant $M_{\rm{BH}}$ measurement,
to within $\sim$0.05 dex uncertainty
\citep[see, e.g.,][]{ben07, par12, bar15}.
\citet{rua15} also find that for their changing-look quasars,
the decrease in luminosity coincides with a broadening of the line widths
to preserve the derived $M_{\rm{BH}}$.
Unfortunately, we cannot test this correlation in this paper,
because we derive the host-galaxy free continuum flux
from a 2D image decomposition of the SDSS images \citep{ben15},
taken at yet a different time from the two sets of spectra.
 We consider this approach superior over spectral decomposition,
  given unknown aperture effects for spectra. (For the same reason, image decomposition
  based on typically an HST image taken at a different time is also used
  for reverberation mapped AGNs \citep[e.g.,][]{ben13}.)
  Moreover, the Keck spectra are not absolutely
  flux calibrated due to typically non-photometric observing conditions.
  However, using the same continuum flux, derived from the 2D image decomposition,
  but taking into account the different width of H$\beta$ between the
  SDSS and Keck spectra, the resulting $M_{\rm{BH}}$ is on average
  0.05$\pm$0.03 dex larger for SDSS spectra. This is small compared to the uncertainty of
  single-epoch measurements of 0.4 dex. (Note that this includes only the 79 objects from \citet{ben15},
  since we do not have continuum measurement for the other objects.
  Also, for obvious reasons, it excludes any object without broad H$\beta$ emission in the Keck spectra.)

The lower left panel of Figure~\ref{fig:ratios} shows that while overall,
the H$\beta_{\rm{narrow}}$/[OIII] flux ratio as measured in the SDSS spectra
is comparable to that in the Keck spectra (on average 1.04$\pm$0.32)
with the majority of the objects falling near the unity line,
the scatter is large due to some extreme outliers.
While the large scatter could partially be due to
the change in broad H$\beta$ (as a consequence of the observed Seyfert-type change)
effecting the narrow H$\beta$ flux for fitting reasons since the lines are blended,
we consider such an effect negligible given the quality of our data and our fitting procedure.
More likely, it might indicate that the NLR emission lines are not constant over
the timescales of these observations but do indeed reverberate, in line with recent studies by \citet[][]{pet13,bar15}.

In the lower right panel of Figure~\ref{fig:ratios}, the broad H$\beta$ to narrow H$\beta$ peak-flux ratio
varies quite a bit between SDSS and Keck spectra (1.26$\pm$0.12). This reflects the change
in the broad H$\beta$ emission line, and, as a consequence, Seyfert type.
On average, the broad H$\beta$ line is stronger
in the SDSS spectra than in the Keck spectra, most likely due the selection bias.
The change in peak-flux ratio can be used as an independent way to classify Seyfert-type transitions,
as discussed in Section~\ref{transitions}.

Reverberation mapping studies have revealed an anti-correlation between broad H$\beta$ width and luminosity
\citep{den09, par12, bar15}, which is attributed to the relation between ionizing flux and the local re-processing efficiency of the BLR gas:
the H$\beta$ re-processing efficiency is greatest in the outer part of the BLR where the flux from the continuum is lower
\citep{kor04, goa14}.  Therefore, higher levels of continuum luminosity lead to an increase in emissivity-weighted BLR radius.  This
so-called breathing effect increases flux for low-velocity line core
relative to high-velocity wings, which makes the line profile narrower.
BLR breathing occurs on short timescales of days to weeks in response to AGN continuum variations.
However, our data do not generally support this special kind of line variability.

Variable accretion and/or variable obscuration are considered
the two main causes for a type change.
As discussed above, to distinguish between them, the
variability in X-ray, UV, and optical is typically studied.  
However, we do not have X-ray data concurring with the SDSS and Keck spectra.
Fig.~\ref{fig:hbeta20}--\ref{fig:hbeta23} show a qualitative comparison between
the change in H$\beta$ and power-law, overlaying the unsubtracted spectra of SDSS and Keck for each object,
scaled to [OIII]. There are 41/102 ($\sim$40\%) objects that have a stronger power-law emission in Keck than SDSS, for
17/102 ($\sim$17\%) it is the other way around, and 44/102 ($\sim$43\%) show no notable change.
The higher fraction of non-stellar (power-law) continuum in the Keck spectra is explainable by
the smaller slit and sharper seeing.

In other words, both the broad H$\beta$ emission line, as well as the power-law continuum,
are varying between the two set of spectra.
However, when attempting to quantify those changes,
we do not find them to be directly correlated.
This is not too surprising, since any variation between broad H$\beta$ emission
and power-law continuum are offset in time depending on the time-lag
of a given object and would not show up at the same time
in a single-epoch spectrum \citep[see e.g., Figs. 7 \& 8 in][]{par12}.

Aperture effects may play a role in our results:
Keck spectra (1\arcsec$\times$1\arcsec~square), given the seeing,
only include the unresolved emission from BLR, AGN power-law continuum, and NLR, while the SDSS spectra (1.5\arcsec~radius circular fiber) may additionally
include more extended NLR flux.
However, any aperture effect 
would artificially boost the ratio of broad-to-narrow flux in Keck spectra (both peak and integrated)
compared to SDSS spectra,
since Keck spectra are restricted to a smaller central area and thus focus on the unresolved emission.
Thus, aperture effects cannot explain the opposite trend that we are observing in the majority of objects,
namely, that we observe less broad H$\beta$ in the Keck spectra.

We note that a few objects (e.g., 0909+1330, 1312+2628, 1708+2153, 2140+0025) show
  significantly stronger blue continuum emission (Fig.~\ref{fig:hbeta20}--\ref{fig:hbeta23}).
  This rise in the blue wavelength range cannot simply be explained by the fact that the Keck spectra
  were not obtained at parallactic angle since this would have the opposite effect.
    Similarly, none of the SDSS spectra were taken at large airmasses,
  with the exception of 2140+0025 that was observed at an airmass of 1.3,
  which could have reduced the blue wavelengths emission artificially for that object in the SDSS spectrum.  
  To further test whether the rise in the blue wavelengths emission in the Keck spectra
  compared to SDSS is an artifact of Keck flux calibration,
  we looked at the spatially-resolved spectra. The rise in the blue continuum is only present
  in the central spectra within the seeing limits, but not in the outer spectra,
  suggesting that it is a real trend.
  A stronger power-law continuum may indeed explain the Seyfert-type change observed for 0909+1330
(from 1.8 in SDSS to 1 in Keck) and 1708+2153 (from 1.5 in SDSS to 1 in Keck).
(Note that the other two objects were classified as Seyfert-1 in both spectra).
  However, as mentioned above, part of this higher fraction of power-law continuum in the Keck spectra
can simply be due to the smaller aperture and sharper seeing compared to SDSS.

\begin{deluxetable*}{lc}
\tabletypesize{\scriptsize}
\tablecolumns{2}
\tablecaption{Comparison between SDSS and Keck}
\tablehead{
  \colhead{Data} &
  \colhead{Average}\\
  \colhead{(1)} & \colhead{(2)}}
\startdata
$\sigma_{\rm{H\beta}}$ & 1.07 $\pm$ 0.03 \\
FHWM$_{\rm{H\beta}}$ & 1.08 $\pm$ 0.03 \\
H$\beta_{\rm{narrow}}$/[OIII] & 1.04 $\pm$ 0.03 \\
H$\beta_{\rm{broad}}$/H$\beta_{\rm{narrow}}$ & 1.26 $\pm$ 0.12
\enddata
\tablecomments{
  Col. (1): Comparison between values derived from the SDSS spectra
  vs. those derived from the Keck spectra for quantities listed in this column.
  Col. (2): Average and scatter.}
\label{statistics}
\end{deluxetable*}

\subsection{Extreme Seyfert-Type Changes}
\label{extreme}
For three objects in the sample, the broad H$\beta$ component was very prominent in the SDSS spectra,
but decreased significantly and  virtually disappeared in the Keck spectra.
All three were re-observed with the 3m Shane telescope
of Lick observatory \citep{sco13}.
Table~\ref{table:extreme} summarizes the Seyfert-type changes of these objects.
Note that all Keck objects classified
conservatively as type-1.9 could be a type-2 object; however, the spectra do not extend to H$\alpha$, so we cannot distinguish between the two. Especially in those cases where the Lick spectrum
reveals a type-2 object, it is likely that the object was also a type-2 in the Keck spectrum.
Also note that for 1423+2720, the SDSS spectrum has a low S/N which makes it difficult
to model the underlying broad H$\beta$ line.

\begin{deluxetable*}{lcccc}
\tabletypesize{\scriptsize}
\tablecolumns{5}
\tablecaption{Extreme Seyfert-Type Changes}
\tablehead{
  \colhead{Object} &
  \colhead{Class.} & \colhead{Class.} & \colhead{Class.} & \colhead{Notes}\\
& \colhead{SDSS} & \colhead{Keck} & \colhead{Lick} \\
  \colhead{(1)} & \colhead{(2)} & \colhead{(3)} & \colhead{(4)} & \colhead{(5)}}\\
\startdata
0847+1824 & 1 & 1.9 & 2 & Off-centered emission in SDSS image\\
          &   &     &   & Type-1 also in literature spectrum taken $\sim$1.8 years before SDSS\\
1038+4658 & 1.5 & 1.9 & 1.9 & Off-centered blob in SDSS image\\
          &     &     &     & \\
1423+2720 & 1.5 & 1.9 & 1.8 & Low S/N in SDSS spectrum\\
\enddata
\tablecomments{
  Col. (1): Object (for more details see Table~\ref{table:observations}).
  Col. (2): Seyfert-type classification based on SDSS spectrum.
Col. (3): Seyfert-type classification based on Keck spectrum.
Col. (4): Seyfert-type classification based on Lick spectrum.
Col. (5): Notes (see text for further discussion).}
\label{table:extreme}
\end{deluxetable*}

The continuum luminosity at 5100\AA~was compared for all objects in the sample in order to determine if there was a correlation between AGN luminosity and strongly variable objects, but  no correlation was found.

For all objects, we carefully searched the literature for other optical spectra.
0847+1824 is the only object for which this search was successful:
It was previously observed on 02-28-2004 (MJD 53063), before the SDSS observation, and from that spectrum, the AGN was a type-1 Seyfert \citep{ho09}.

Apart from variable accretion and/or obscuration, we briefly discuss a few other scenarios which could
mimic changes in Seyfert type.

(i) Telescope offset: First of all, we note that apparent Seyfert-type changes
could be caused by a slight mis-pointing of the Keck telescope, missing the AGN
core and therefore the (bulk of the) broad emission lines.
In all cases, the telescope was pointed at the center of the galaxy (as verified by
  guide star images),
assuming that the AGN resides there.
For a couple of extreme objects,
the AGN might actually be offset from the center, as evidenced by the SDSS images;
we will discuss them below.
Note that for all objects for which the AGN core and BLR coincide with the 
  center of the galaxy, the possibility of missing the BLR emission due to a telescope offset
  is negligible, given the seeing ($\sim$1\arcsec~for the Keck observations) and the slit width used
  (1\arcsec, matching the typical seeing), as verified by standard-star observations. Keck telescope
guiding is also much more accurate than 1\arcsec.

(ii) Galaxy mergers and/or recoiling SMBH: In the course of ongoing
galaxy mergers, the AGN can appear offset from the apparent center
of the merging system. It is therefore possible that an off-center AGN was
captured by the wider, circular SDSS aperture, but missed
during subsequent observation by the narrower Keck slit, aligned along
the major axis of the galaxy and centered on the galactic nucleus.
Alternatively, a rare gravitational wave recoil following the final
coalescence of two SMBHs in a merger, can remove the newly formed
single SMBH from the center of its host galaxy
\citep[e.g., see the review by][]{spe15}.
The accretion disk and BLR would remain bound, and a recoiling SMBH would
therefore appear as an AGN offset from the core of its host galaxy
\citep[][]{kom12},
again leading to the possibility of missing the (bulk of the) BLR
in the more narrow (and rectangular) Keck slit.

With the exception of a few cases
\citep[4/102, see][]{ben15};
there are no signs for merger activity in the sample.
However, we caution that merger signatures such as faint tidal
tails might easily be missed in the low S/N SDSS images.
Objects 0847+1824 and 1038+4658 show extended emission in the
SDSS multi-color images, offset from the galaxy center
(Fig.~\ref{appendix:sdss}), which might indicate the
presence of an ongoing merger. While we cannot exclude the
possibility that these are off-center AGNs,
it is statistically unlikely to have
off-centered AGNs in such a small sample,
and follow-up spectroscopy
would be needed to test such a scenario further. We note in passing
that 0847+1824 seems to show a small kinematic offset between its narrow
and broad H$\beta$ line (with the broad H$\beta$ line being blue-shifted by
$\sim$100km\,s$^{-1}$), which is, however, most likely mimicked by
the asymmetric broad line profile.

(iii) Supernovae: A type IIn supernova has many of the
same spectral features as a Seyfert galaxy \citep[][]{fil97}.
A nuclear supernova could have therefore mimicked the presence
of an AGN. However, supernova spectra including the narrow emission
lines evolve rapidly, and we do not see any other signs of dramatic
changes in the continuum and narrow emission lines.

(iv) Stellar tidal disruption event: Stars can be tidally disrupted and
accreted by SMBHs, producing a luminous accretion flare
\citep[e.g.,][]{ree88}.
If these occur in a gas rich environment, broad and narrow emission lines
can be temporarily excited. While a few candidate events for this process
have been identified recently from SDSS \citep[e.g.,][]{kom08},
these events are rare, and are unlikely to occur in our small sample.
In particular, we have checked the long-term Catalina lightcurves of all
three sources, and none shows the characteristic lightcurve decline expected
for a typical tidal disruption event.

We are left with mild changes in accretion or extinction as most likely
explanation for the three changing-look AGN in our sample. Future spectroscopic
monitoring of emission-line and continuum changes will enable us to distinguish
between both possibilities.
(Note that a difference in aperture between Keck and SDSS cannot explain the extreme Seyfert-type changes we observe in these three objects
  since it would have the opposite effect.)

\subsection{Comparison with the Catalina Sky Survey}
To further shed light on the causes for the observed variability,
we considered the optical light curves in 
the Catalina Sky Survey (CSS) \citep[][]{dra09}.
With the exception of 1104+4334 and 1206+4244,
all objects in our sample are in the CSS archive. 
(Note that 1605+3305 is in the archive, but it does not have a light curve available.)
For most objects, 
CSS light curves begin after the SDSS observations, but extend past the time of the Keck and Lick observations;
the light curves start and end at approximately $\sim$53500$-$56500 MJD ($\sim$3000 days).
The light curves of seven objects
(0310-0049, 0904+5536, 1147+0902, 1355+3834, 1434+4839, 1535+5754, 1557+0830)\footnote{Note that 0301+0110 shows a highly variable light-curve, but upon further inspection, it becomes clear that that is an artifact of a bright nearby star.}
reveal large variability
($\sim$0.5-1mag in one object)
over time-scales from days to months to years.
None of these objects are amongst our extreme subset of objects (see Table~\ref{table:observations} for Seyfert types).

Since CSS photometry is aperture based
\citep{dra09}, we cannot exclude that a variable seeing
can mimic variability since more or less of the host galaxy would be included
in an aperture centered on the AGN. However, especially the extreme variability
in these seven objects is unlikely to be purely a seeing effect.
Independent analysis, which is beyond the scope of this paper, is needed to further confirm the observed variability.

\section{Summary}
\label{summary}
In this paper, we study the broad H$\beta$ emission-line variability in a sample of 102 local Seyfert-1 galaxies, selected from SDSS and re-observed 3-9 (on average 6.4$\pm$1.8) years later with LRIS on the 10m Keck-I telescope.

In the $3-9$ year time-frame between observations,
67/102 ($\sim$66\%) objects show at least some form
  of variability of either width and/or strength of the broad H$\beta$ line.
  For 39/102 ($\sim$38\%) objects, this variability is significant enough
  to result in a change in Seyfert type, following the standard
  Seyfert classification scheme.
  There is no correlation between the time between observations and the
  degree of the observed Seyfert-type transition, implying that the
  transitions happen on shorter time scales.
  Short-time variability on the scale of days and weeks
  is known for low-mass AGNs from reverberation mapping.
  Almost all objects (99/102) were observed as part of the CSS
  with 7/102 ($\sim$7\%) displaying significant variability 
 on time scales of days to weeks.  
  
Three ($\sim$3\%) objects are extreme cases for which
  the broad H$\beta$ component almost completely disappears.
  We discuss possible origins for these transitions.
  For two of these objects (0847+1824 and 1038+4658),
  SDSS images reveal extended emission off-centered from the
  galaxy center that could have been included in the SDSS spectra,
  but missed by the smaller-area Keck slit centered on the galaxy.

  The study presented here is the first to provide statistical information on the frequency
  and strength of H$\beta$ line variability in a sample of low-redshift Seyfert galaxies.
  
  \acknowledgments
We thank the anonymous referee for valuable comments
helping to improve the paper.
We thank Aaron Barth and Bernd Husemann for helpful discussions, Luis Ho for providing additional data, and William C. Keel for
data reduction of the Lick spectra.
JNR, MC, and VNB gratefully acknowledge
assistance from a National Science Foundation (NSF) Research at Undergraduate Institutions (RUI) grant AST-1312296.  Note that findings and conclusions do not necessarily represent views of the NSF.
VNB, BS, and SK would like to thank the Kavli Institute for Theoretical 
Physics (Santa Barbara) for their hospitality and support; 
the KITP is supported by NSF Grant No. NSF PHY11-25915.
D.P. acknowledges support through the EACOA Fellowship from The East Asian Core Observatories Association, which consists of the National Astronomical Observatories, Chinese Academy of Science (NAOC), the National Astronomical Observatory of Japan (NAOJ), Korean Astronomy and Space Science Institute (KASI), and Academia Sinica Institute of Astronomy and Astrophysics (ASIAA).
This research has made use of the Dirac computer cluster at Cal Poly, maintained by Dr. Brian Granger and Dr. Ashley Ringer McDonald.  Data presented in this thesis were obtained at the W. M. Keck Observatory, which is operated as a scientific partnership among Caltech, the University of California, and NASA.  The Observatory was made possible by the generous financial support of the W. M. Keck Foundation.  The authors recognize and acknowledge the very significant cultural role and reverence that the summit of Mauna Kea has always had within the indigenous Hawaiian community.  We are most fortunate to have the opportunity to conduct observations from this mountain.  This research has made use of the public archive of the Sloan Digital Sky Survey (SDSS) and the NASA/IPAC Extragalactic Database (NED) which is operated by the Jet Propulsion Laboratory, California Institute of Technology, under contract with the National Aeronautics and Space Administration.  The CSS survey is funded by the National Aeronautics and Space Administration under Grant No. NNG05GF22G issued through the Science Mission Directorate Near-Earth Objects Observations Program.  The CRTS survey is supported by the U.S.~National Science Foundation under grants AST-0909182 and AST-1313422.

Facilities: \facility{Keck: I (LRIS), Lick: Shane 3-m Kast spectrograph}

\appendix

\section{Table of Observations and Seyfert-Type Classification}
\label{appendix:observations}
  \LongTables
  \begin{deluxetable*}{lccccccccccc}
  \tabletypesize{\scriptsize} 
\tablecolumns{12}
\tablecaption{Observations and Seyfert-Type Classification}
\tablehead{
  \colhead{Object} &
  \colhead{R.A.} &
  \colhead{Dec.} &
  \colhead{{\it z}} &
  \colhead{Date} &
  \colhead{Class.} &
  \colhead{Date} &
  \colhead{Class.} &
  \colhead{Exp. time} &
  \colhead{Diff.} &
  \colhead{Date} &
  \colhead{Class.}\\
  & \colhead{(J2000)} & \colhead{(J2000)} &  & \colhead{SDSS} & \colhead{SDSS} & \colhead{Keck} &\colhead{Keck} & \colhead{Keck [s]}  & \colhead{years} & \colhead{Lick} & \colhead{Lick}\\
  \colhead{(1)} & \colhead{(2)} &
 \colhead{(3)} & \colhead{(4)} &
 \colhead{(5)} & \colhead{(6)} & 
 \colhead{(7)} & \colhead{(8)} &
 \colhead{(9)} & \colhead{(10)} &
 \colhead{(11)} & \colhead{(12)}}
 \startdata
 0013$-$0951 & 00 13 35.38 & $-$09 51 20.9 & 0.062 & 08-17-2001 & 1 & 09-20-2009 & 1.5 & 600 & 8.09\\
 & & & &(52138)& &(55094)& &\\ 
 0026+0009 & 00 26 21.29 & +00 09 14.9 & 0.060 & 08-26-2000 & 1 & 09-20-2009 & 1 & 1600 & 9.07\\
 & & & &(51782)& &(55094)& & & \\
 0038+0034 & 00 38 47.96 & +00 34 57.5 & 0.081 & 09-06-2000 & 1 & 09-20-2009 & 1.5 & 600 & 9.04\\
 & & & &(51793)& &(55094)& & & \\ 
 0109+0059 & 01 09 39.01 & +00 59 50.4 & 0.093 & 09-07-2000 & 1.5 & 09-20-2009 & 1.5 & 600 & 9.04\\
  & & & &(51794)& &(55094)& & & \\
0121$-$0102 & 01 21 59.81 & $-$01 02 24.4 & 0.054 & 09-02-2000 & 1.5 & 01-21-2009 & 1 & 1200 & 8.39\\
 & & & &(51789)& &(54852)& & & \\
0150+0057 & 01 50 16.43 & +00 57 01.9 & 0.085 & 09-06-2000 & 1 & 09-20-2009 & 1 &600 & 9.04\\
 & & & &(51793)& &(55094)& & & \\
0206$-$0017 & 02 06 15.98 & $-$00 17 29.1 & 0.043 & 09-25-2000 & 1 & 01-22-2009 & 1 &1200 & 8.33\\
 & & & &(51812)& &(54853)& & & \\
0212+1406 & 02 12 57.59 & +14 06 10.0 & 0.062 & 12-05-2000 & 1 & 09-20-2009 & 1 &600 & 8.79\\
 & & & &(51883)& &(55094)& & & \\
0301+0110 & 03 01 24.26 & +01 10 22.5 & 0.072 & 09-30-2000 & 1.5 & 09-20-2009 & 1.5 &600 & 8.97\\
 & & & &(51817)& &(55094)& & & \\
0301+0115 & 03 01 44.19 & +01 15 30.8 & 0.075 & 09-30-2000 & 1 & 09-20-2009 & 1 &600 & 8.97\\
 & & & &(51817)& &(55094)& & & \\
0310$-$0049 & 03 10 27.82 & $-$00 49 50.7 & 0.080 & 12-15-2001 & 1 & 09-20-2009 & 1 & 600 & 7.76\\
 & & & &(52258)& &(55094)& & & \\
 0336$-$0706 & 03 36 02.09 & $-$07 06 17.1 & 0.097 & 12-31-2000 & 1.8 & 09-20-2009 & 1.8 & 2400 & 8.72\\
  & & & &(51909)& &(55094)& & & \\
0353$-$0623 & 03 53 01.02 & $-$06 23 26.3 & 0.076 & 12-30-2000 & 1.8 & 01-22-2009 & 1 & 1200 & 8.06\\
 & & & &(51908)&  &(54853)& & & \\
0731+4522 & 07 31 26.68 & +45 22 17.4 & 0.092 & 11-05-2004 & 1.5 & 09-20-2009 & 1.5  & 600 & 4.87\\
 & & & &(53314)& & (55094)& & & \\
0735+3752 & 07 35 21.19 & +37 52 01.9 & 0.096 & 11-29-2000 & 1.5 & 09-20-2009 & 1.8 &600 & 8.81\\
 & & & &(51877)& &(55094)& & & \\
0737+4244 & 07 37 03.28 & +42 44 14.6 & 0.088 & 01-31-2004 & 1.5 & 09-20-2009 & 1.5 & 600 & 5.64\\
 & & & &(53035)& &(55094)& & & \\
0802+3104 & 08 02 43.40 & +31 04 03.3 & 0.041 & 01-02-2003 & 1 & 01-21-2009 & 1 & 1200 & 5.97\\
 & & & &(52641)& &(54852)& & & \\
0811+1739 & 08 11 10.28 & +17 39 43.9 & 0.065 & 12-18-2004 & 1.5 & 03-15-2010 & 1 & 2700 & 5.24\\
 & & & &(53357)& &(55270)& & & \\
0813+4608 & 08 13 19.34 & +46 08 49.5 & 0.054 & 11-29-2000 & 1.8 & 01-14-2010 & 1 & 1200 & 9.13\\
 & & & &(51877)& &(55210)& & & \\
0831+0521 & 08 31 07.62 & +05 21 05.9 & 0.035 & 01-07-2003 & 1.8 & 03-15-2010 & 1.9 & 600 & 7.18\\
 & & & &(52646)& &(55270)& & & \\
0845+3409 & 08 45 56.67 & +34 09 36.3 & 0.066 & 02-02-2003 & 1.5 & 03-14-2010 & 1.5 &3600 & 7.11\\
 & & & &(52672)& &(55269)& & & \\
0846+2522 & 08 46 54.09 & +25 22 12.3 & 0.051 & 12-19-2004 & 1.5 & 01-22-2009 & 1.5 & 1200 & 4.09\\
 & & & &(53358)& &(54853)& & & \\
0847+1824 & 08 47 48.28 & +18 24 39.9 & 0.085 & 12-07-2005 & 1 & 01-21-2009 & 1.9 & 1200 & 3.10 & 01-15-2013 & 2\\
 & & & &(53711)& &(54852)& & & &(56307) & \\
0854+1741 & 08 54 39.25 & +17 41 22.5 & 0.065 & 12-25-2005 & 1.5 & 03-15-2010 & 1 & 600 & 4.22\\
 & & & &(53729)& &(55270)& & & \\
0857+0528 & 08 57 37.77 & +05 28 21.3 & 0.059 & 01-31-2003 & 1 & 01-15-2010 & 1 & 600 & 6.96\\
 & & & &(52670)& &(55211)& & & \\
0904+5536 & 09 04 36.95 & +55 36 02.5 & 0.037 & 12-30-2000 & 1.5 & 03-14-2010 & 1.5 &600 & 9.20\\
 & & & &(51908)& &(55269)& & & \\
0909+1330 & 09 09 02.35 & +13 30 19.4 & 0.051 & 04-01-2006 & 1.8 & 01-14-2010 & 1 & 600 & 3.79\\
 & & & &(53826)& &(55210)& & & \\
0921+1017 & 09 21 15.55 & +10 17 40.9 & 0.039 & 02-15-2004 & 1.8 & 01-14-2010 & 1.8 & 700 & 5.91\\
 & & & &(53050)& &(55210)& & & \\
0923+2254 & 09 23 43.00 & +22 54 32.7 & 0.033 & 12-23-2005 & 1 & 01-15-2010 & 1 & 600 & 4.06\\
 & & & &(53727)& &(55211)& & & \\
0923+2946 & 09 23 19.73 & +29 46 09.1 & 0.063 & 01-19-2005 & 1.8 & 01-15-2010 & 1.8 & 600 & 4.99\\
 & & & &(53389)& &(55211)& & & \\
0927+2301 & 09 27 18.51 & +23 01 12.3 & 0.026 & 12-26-2005 & 1.5 & 01-15-2010 & 1.5 & 600 & 4.05\\
 & & & &(53730)& &(55211)& & & \\
0932+0233 & 09 32 40.55 & +02 33 32.6 & 0.057 & 02-25-2001 & 1.8 & 01-14-2010 & 1.5 & 600 & 8.88\\
 & & & &(51965)& &(55210)& & & \\
0932+0405 & 09 32 59.60 & +04 05 06.0 & 0.059 & 12-21-2001 & 1.8 & 01-14-2010 & 1.9 & 600 & 8.07 & 03-11-2013 & 1.9\\
& & & &(52264)& &(55210)& & & & (56362) & \\
0936+1014 & 09 36 41.08 & +10 14 15.7 & 0.060 & 12-20-2003 & 1.5 & 03-15-2010 & 1 & 3600 & 6.23\\
 & & & &(52993)& &(55270)& & & \\
0938+0743 & 09 38 12.27 & +07 43 40.0 & 0.022 & 04-04-2003 & 1 & 01-14-2010 & 1.8 & 600 & 6.78 & 01-15-2013 & 1.8\\
 & & & &(52733)& &(55210)& & & & (56307)\\
0948+4030 & 09 48 38.43 & +40 30 43.5 & 0.047 & 03-11-2003 & 1 & 01-15-2010 & 1.8 & 900 & 6.85\\
 & & & &(52709)& &(55211)& & & \\
1002+2648 & 10 02 18.79 & +26 48 05.7 & 0.052 & 01-22-2006 & 1.8 & 01-15-2010 & 1.9 & 600 & 3.98\\
 & & & &(53757)& &(55211)& & & \\
1029+1408 & 10 29 25.73 & +14 08 23.2 & 0.061 & 03-11-2004 & 1.5 & 01-15-2010 & 1.5 & 600 & 5.85\\
 & & & &(53075)& &(55211)& & & \\
1029+2728 & 10 29 01.63 & +27 28 51.2 & 0.038 & 02-28-2006 & 1.8 & 01-15-2010 & 1.8 & 600 & 3.88\\
 & & & &(53794)& &(55211)& & & \\
1029+4019 & 10 29 46.80 & +40 19 13.8 & 0.067 & 01-29-2004 & 1.5 & 01-14-2010 & 1.5 & 600 & 5.96\\
 & & & &(53033)& &(55210)& & & \\
1038+4658 & 10 38 33.42 & +46 58 06.6 & 0.063 & 12-12-2002 & 1.5 & 01-14-2010 & 1.9 & 600 & 7.09 & 01-17-2013 & 1.9\\
 & & & &(52620)& &(55210)& & & & (56309)\\
1042+0414 & 10 42 52.94 & +04 14 41.1 & 0.052 & 03-06-2002 & 1.5 & 04-16-2009 & 1.5 & 1200 & 7.11\\
 & & & &(52339)& &(54937)& & & \\
1043+1105 & 10 43 26.47 & +11 05 24.3 & 0.048 & 04-20-2004 & 1.8 & 04-16-2009 & 1.8 & 600 & 4.99\\
 & & & &(53115)& &(54937)& & & \\
1049+2451 & 10 49 25.39 & +24 51 23.7 & 0.055 & 02-26-2006 & 1 & 04-16-2009 & 1 & 600 & 3.13\\
 & & & &(53792)& &(54937)& & & \\
1058+5259 & 10 58 28.76 & +52 59 29.0 & 0.068 & 01-13-2003 & 1.5  & 01-14-2010 & 1.5 & 600 & 7.00\\
 & & & &(52652)& &(55210)& & & \\
1101+1102 & 11 01 01.78 & +11 02 48.8 & 0.036 & 04-24-2004 & 1.5 & 04-16-2009 & 1.8 & 600 & 4.98\\
 & & & &(53119)& &(54937)& & & \\
1104+4334 & 11 04 56.03 & +43 34 09.1 & 0.049 & 02-18-2004 & 1.8 & 01-14-2010 & 1.5 & 600 & 5.91\\
 & & & &(53053)& &(55210)& & & \\
1110+1136 & 11 10 45.97 & +11 36 41.7 & 0.042 & 03-14-2004 & 1.5 & 03-15-2010 & 1 & 3600 & 6.00\\
 & & & &(53078)& &(55270)& & & \\
1116+4123 & 11 16 07.65 & +41 23 53.2 & 0.021 & 12-30-2003 & 1.8 & 04-15-2009 & 1.8 & 850 & 5.29\\
 & & & &(53003)& &(54936)& & & \\
1118+2827 & 11 18 53.02 & +28 27 57.6 & 0.060 & 02-27-2006 & 1.8 & 01-15-2010 & 1.9 & 900 & 3.88 & 03-11-2013 & 1.8\\
 & & & &(53793)& &(55211)& & & & (56362)\\
1132+1017 & 11 32 49.28 & +10 17 47.4 & 0.044 & 05-22-2003 & 1.5 & 01-15-2010 & 1 & 600 & 6.65\\
 & & & &(52781)& &(55211)& & & \\
1137+4826 & 11 37 04.17 & +48 26 59.2 & 0.054 & 01-03-2003 & 1.5 & 01-14-2010 & 1.5 & 600 & 7.03\\
 & & & &(52642)& &(55210)& & & \\
1139+5911 & 11 39 08.95 & +59 11 54.6 & 0.061 & 05-15-2002 & 1 & 01-14-2010 & 1 & 600 & 7.67\\
 & & & &(52409)& &(55210)& & & \\
1140+2307 & 11 40 54.09 & +23 07 44.4 & 0.035 & 05-21-2006 & 1.8 & 01-15-2010 & 1.8 & 1200 & 3.66 & 01-13-2013 & 2\\
 & & & &(53876)& &(55211)& & & & (56305)\\
1143+5941 & 11 43 44.30 & +59 41 12.4 & 0.063 & 05-17-2002 & 1.5 & 03-14-2010 & 1 & 3000 & 7.82\\
 & & & &(52411)& &(55269)& & & \\
1144+3653 & 11 44 29.88 & +36 53 08.5 & 0.038 & 03-13-2005 & 1 & 04-16-2009 & 1 & 600 & 4.09\\
 & & & &(53442)& &(54937)& & & \\
1145+5547 & 11 45 45.18 & +55 47 59.6 & 0.053 & 04-30-2003 & 1 & 03-14-2010 & 1 & 3600 & 6.87\\
 & & & &(52759)& &(55269)& & & \\
1147+0902 & 11 47 55.08 & +09 02 28.8 & 0.069 & 05-01-2003 & 1.5 & 01-15-2010 & 1.5 & 600 & 6.71\\
 & & & &(52760)& &(55211)& & & \\
1205+4959 & 12 05 56.01 & +49 59 56.4 & 0.063 & 06-17-2002 & 1.8 & 01-14-2010 & 1.8 & 600 & 7.58\\
 & & & &(52442)& &(55210)& & & \\
1206+4244 & 12 06 26.29 & +42 44 26.1 & 0.052 & 04-25-2004 & 1 & 03-14-2010 & 1 & 1100 & 5.88\\
 & & & &(53120)& &(55269)& & & \\
1210+3820 & 12 10 44.27 & +38 20 10.3 & 0.023 & 04-13-2005 & 1.5 & 04-16-2009 & 1.5 & 600 & 4.01\\
 & & & &(53473)& &(54937)& & & \\
1216+5049 & 12 16 07.09 & +50 49 30.0 & 0.031 & 05-19-2002 & 1.8 & 03-14-2010 & 1.8 & 900 & 6.82\\
 & & & &(52413)& &(55269)& & & \\
1223+0240 & 12 23 24.14 & +02 40 44.4 & 0.024 & 01-09-2002 & 1 & 03-15-2010 & 1 & 600 & 8.18\\
 & & & &(52283)& &(55270)& & & \\
1228+0951 & 12 28 11.41 & +09 51 26.7 & 0.064 & 04-02-2003 & 1.8 & 03-15-2010 & 1.8 & 600 & 6.95\\
 & & & &(52731)& &(55270)& & & \\
1231+4504 & 12 31 52.04 & +45 04 42.9 & 0.062 & 02-27-2004 & 1.5 & 01-15-2010 & 1.5 & 1200 & 5.88\\
 & & & &(53062)& &(55211)& & & \\
1241+3722 & 12 41 29.42 & +37 22 01.9 & 0.063 & 04-02-2006 & 1.5 & 01-15-2010 & 1.5 & 800 & 3.79\\
 & & & &(53827)& &(55211)& & & \\
1246+5134 & 12 46 38.74 & +51 34 55.9 & 0.067 & 04-15-2002 & 1.8 & 01-15-2010 & 1.5 & 600 & 7.75\\
 & & & &(52379)& &(55211)& & & \\
1250$-$0249 & 12 50 42.44 & $-$02 49 31.5 & 0.047 & 03-29-2001 & 1.5 & 04-16-2009 & 1.8 & 1200 & 8.05\\
 & & & &(51997)& &(54937)& & &  \\
1306+4552 & 13 06 19.83 & +45 52 24.2 & 0.051 & 04-22-2004 & 1 & 03-14-2010 & 1.5 & 3600 & 5.89\\
 & & & &(53117)& & (55269)& & & \\
1307+0952 & 13 07 21.93 & +09 52 09.3 & 0.049 & 05-29-2006 & 1.8 & 03-15-2010 & 1.5 & 2400 & 3.79\\
 & & & &(53884)& &(55270)& & & \\
1312+2628 & 13 12 59.59 & +26 28 24.0 & 0.060 & 02-28-2006 & 1 & 03-14-2010 & 1 & 2700 & 4.04\\
 & & & &(53794)& &(55269)& & & \\
1313+3653 & 13 13 48.96 & +36 53 57.9 & 0.067 & 03-21-2006 & 1.8 & 03-14-2010 & 1.8 & 600 & 3.98\\
 & & & &(53815)& &(55269)& & & \\
1323+2701 & 13 23 10.39 & +27 01 40.4 & 0.056 & 03-01-2006 & 1.8 & 04-16-2009 & 1.8 & 700 & 3.13\\
 & & & &(53795)& &(54937)& & & \\
1353+3951 & 13 53 45.93 & +39 51 01.6 & 0.063 & 02-26-2004 & 1.8 & 03-14-2010 & 1.9 & 600 & 6.05 & 03-12-2013 & 1.8\\
 & & & &(53061)& &(55269)& & & & (56363)\\
1355+3834 & 13 55 53.52 & +38 34 28.5 & 0.050 & 03-31-2005 & 1.8 & 04-16-2009 & 1.8 & 300 & 4.04\\
 & & & &(53460)& &(54937)& & & \\
1405$-$0259 & 14 05 14.86 & $-$02 59 01.2 & 0.054 & 06-18-2002 & 1 & 04-16-2009 & 1 & 1600 & 6.83\\
 & & & &(52443)& &(54937)& & & \\
1416+0317 & 14 16 30.82 & +01 37 07.9 & 0.054 & 03-26-2001 & 1.5 & 03-15-2010 & 1.8 & 2700 & 8.97\\
 & & & &(51994)& &(55270)& & & \\
1419+0754 & 14 19 08.30 & +07 54 49.6 & 0.056 & 06-12-2005 & 1.8 & 04-16-2009 & 1.8 & 900 & 3.84\\
 & & & &(53533)& &(54937)& & & \\
1423+2720 & 14 23 38.43 & +27 20 09.7 & 0.064 & 05-26-2006 & 1.5 & 03-14-2010 & 1.9 & 1200 & 3.80 & 03-12-2013 & 1.8\\
 & & & &(53881)& &(55269)& & &  & (56363)\\
1434+4839 & 14 34 52.45 & +48 39 42.8 & 0.037 & 04-04-2003 & 1 & 04-16-2009 & 1.5 & 600 & 6.03\\
 & & & &(52733)& &(54937)& & & \\
1505+0342 & 15 05 56.55 & +03 42 26.3 & 0.036 & 05-26-2001 & 1.5 & 03-15-2010 & 1.5 & 1200 & 8.80\\
 & & & &(52055)& &(55270)& & & \\
1535+5754 & 15 35 52.40 & +57 54 09.3 & 0.030 & 03-14-2002 & 1 & 04-15-2009 & 1 & 1200 & 7.09\\
 & & & &(52347)& &(54936)& & & \\
1543+3631 & 15 43 51.49 & +36 31 36.7 & 0.067 & 08-24-2003 & 1 & 03-15-2010 & 1.5 & 1200 & 6.56\\
 & & & &(52875)& &(55270)& & & \\
1545+1709 & 15 45 07.53 & +17 09 51.1 & 0.048 & 06-03-2006 & 1.8 & 04-15-2009 & 1 &  1200 & 2.57\\
 & & & &(53889)& &(54936)& & & \\
1554+3238 & 15 54 17.42 & +32 38 37.6 & 0.048 & 07-05-2003 & 1.5 & 04-15-2009 & 1.5 & 1200 & 5.78\\
 & & & &(52825)& &(54936)& & & \\
1557+0830 & 15 57 33.13 & +08 30 42.9 & 0.047 & 08-11-2004 & 1.5 & 04-15-2009 & 1.5 & 1200 & 4.68\\
 & & & &(53228)& &(54936)& & & \\
1605+3305 & 16 05 02.46 & +33 05 44.8 & 0.053 & 05-17-2004 & 1 & 04-15-2009 & 1 & 1200 & 4.91\\
 & & & &(53142)& &(54936)& & & \\
1606+3324 & 16 06 55.94 & +33 24 00.3 & 0.059 & 05-19-2004 & 1.5 & 04-15-2009 & 1.5 & 1200 & 4.91\\
 & & & &(53144)& &(54936)& & & \\
1611+5211 & 16 11 56.30 & +52 11 16.8 & 0.041 & 05-22-2001 & 1.5 & 04-15-2009 & 1.8 & 1200 & 7.95\\
 & & & &(52051)& &(54936)& & & \\
1636+4202 & 16 36 31.28 & +42 02 42.5 & 0.061 & 06-23-2001 & 1 & 03-14-2010 & 1 & 1200 & 8.72\\
 & & & &(52083)& &(55269)& & & \\
1647+4442 & 16 47 21.47 & +44 42 09.7 & 0.025 & 05-22-2001 & 1.8 & 03-14-2010 & 1.5 & 4200 & 8.81\\
 & & & &(52051)& &(55269)& & & \\
1655+2014 & 16 55 14.21 & +20 14 42.0 & 0.084 & 06-13-2004 & 1.8 & 09-20-2009 & 1.8 & 600 & 5.27\\
 & & & &(53169)& &(55094)& & & \\
1708+2153 & 17 08 59.15 & +21 53 08.1 & 0.072 & 06-21-2004 & 1.5 & 09-20-2009 & 1 & 600 & 5.25\\
 & & & &(53177)& &(55094)& & & \\
2116+1102 & 21 16 46.33 & +11 02 37.3 & 0.081 & 07-13-2002 & 1.8 & 09-20-2009 & 1.8 & 700 & 7.19\\
 & & & &(52468)& &(55094)& & & \\
2140+0025 & 21 40 54.55 & +00 25 38.2 & 0.084 & 07-10-2002 & 1 & 09-20-2009 & 1 & 600 & 7.20\\
 & & & &(52465)& &(55094)& & & \\
2215$-$0036 & 22 15 42.29 & $-$00 36 09.6 & 0.099 & 09-04-2000 & 1 & 09-20-2009 & 1 & 600 & 9.04\\
 & & & &(51791)& &(55094)& & & \\
2221$-$0906 & 22 21 10.83 & $-$09 06 22.0 & 0.091 & 10-21-2001 & 1 & 09-20-2009 & 1 & 600 & 7.92\\
 & & & &(52203)& &(55094)& & & \\
2222$-$0819 & 22 22 46.61 & $-$08 19 43.9 & 0.082 & 10-24-2001 & 1.5 & 09-20-2009 & 1.5 & 700 & 7.91\\
 & & & &(52206)& &(55094)& & & \\
2233+1312 & 22 33 38.42 & +13 12 43.5 & 0.093 & 09-04-2002 & 1 & 09-20-2009 & 1.5 & 800 & 7.04\\
 & & & &(52521)& &(55094)& & & \\
2254+0046 & 22 54 52.24 & +00 46 31.4 & 0.091 & 09-02-2000 & 1 & 09-20-2009 & 1 & 600 & 9.05\\
 & & & &(51789)& &(55094)& & & \\
2327+1524 & 23 27 21.97 & +15 24 37.4 & 0.046 & 11-25-2001 & 1.5 & 09-20-2009 & 1.8 & 600 & 7.82\\
 & & & &(52238)& &(55094)& & & \\
2351+1552 & 23 51 28.75 & +15 52 59.1 & 0.096 & 11-13-2001 & 1.8 & 09-20-2009 & 1.8 & 600 & 7.85\\
& & & &(52226)& &(55094)& & & 
\enddata
\tablecomments{Col. (1): Target ID based on R.A. and Dec. used throughout the text.  Col. (2): Right ascension.  Col. (3): Declination.  Col. (4): Redshift from SDSS-DR7.  Col. (5):
  Date SDSS spectrum was taken (with Modified Julian Date (MJD) in brackets).  Col. (6): Seyfert-type classification based on SDSS spectrum. Col. (7): Date Keck spectrum was taken (with MJD in brackets).  Col. (8): Seyfert-type classification based on Keck spectrum.  Note that because Keck spectra do not extend to H$\alpha$, we cannot differentiate between type-1.9 and type-2, and conservatively classify these objects in question as type-1.9. Col. (9): Exposure time of Keck observations in seconds.
Note that the exposure time for all SDSS spectra is 54 seconds.
  Col. (10): Time between SDSS and Keck observations in years.  Col. (11): Date Lick spectrum was taken (with MJD in brackets). Col. (12): Seyfert-type classification based on Lick spectrum.}
\label{table:observations}
\end{deluxetable*}

\LongTables
\begin{deluxetable*}{lcccccccccccc}
  \tabletypesize{\tiny}
\tablecolumns{13}
\tablecaption{Results from Spectral Fitting}
\tablehead{
  \colhead{Object} &
  \colhead{$\sigma_{H\beta}$} &
  \colhead{$FWHM_{H\beta}$} &
  \colhead{$\frac{H\beta narrow}{[OIII]}$ flux} &
  \colhead{H$\beta \frac{broad}{narrow}$ flux} &
  \colhead{H$\beta \frac{broad}{narrow}$ peak flux} &
  \colhead{FeII} &
  \colhead{$\sigma_{H\beta}$} &
  \colhead{FWHM$_{H\beta}$} &
  \colhead{$\frac{H\beta narrow}{[OIII]}$ flux} &
  \colhead{H$\beta \frac{broad}{narrow}$ flux} &
  \colhead{H$\beta \frac{broad}{narrow}$ peak flux} &
  \colhead{FeII} \\
  & \colhead{SDSS} & \colhead{SDSS} & \colhead{SDSS} & \colhead{SDSS} & \colhead{SDSS} & \colhead{SDSS}& \colhead{Keck} & \colhead{Keck} & \colhead{Keck} & \colhead{Keck} & \colhead{Keck} & \colhead{Keck} \\
  & \colhead{[km s$^{-1}$]} & \colhead{[km s$^{-1}$]} & & & & & \colhead{[km s$^{-1}$]} & \colhead{[km s$^{-1}$]} \\
\colhead{(1)} & \colhead{(2)} &
 \colhead{(3)} & \colhead{(4)} &
 \colhead{(5)} & \colhead{(6)} & 
 \colhead{(7)} & \colhead{(8)} &
 \colhead{(9)} & \colhead{(10)} &
 \colhead{(11)} & \colhead{(12)} & \colhead{(13)}}
\startdata
0013$-$0951 & 1783$\pm$135 & 3462$\pm$205 & 0.13$\pm$0.02 & 65$\pm$10 & 8$\pm$1 & Y & 2111$\pm$211 & 4275$\pm$594 & 0.17$\pm$0.02 & 20$\pm$2 & 2.05$\pm$0.17 & Y\\
0026+0009 & 964$\pm$125 & 2276$\pm$293 & 0.18$\pm$0.05 & 14$\pm$4 & 2.7$\pm$0.7 & N & 1527$\pm$227 & 2920$\pm$765 & 0.26$\pm$0.06 & 10$\pm$2 & 1.4$\pm$0.3 & Y \\
0038+0034 & 2759$\pm$133 & 7371$\pm$321 & 0.08$\pm$0.01 & 23.2$\pm$1.8 & 1.54$\pm$0.09 & N & 3328$\pm$211 & 7223$\pm$593 & 0.09$\pm$0.01 & 10$\pm$1 & 0.63$\pm$0.04 & Y \\
0109+0059 & 1677$\pm$230 & 3392$\pm$282 & 0.15$\pm$0.02 & 4.6$\pm$0.8 & 0.70$\pm$0.09 & Y & 1797$\pm$268 & 3377$\pm$355 & 0.11$\pm$0.01 & 4.1$\pm$0.8 & 0.61$\pm$0.06 & Y\\
0121$-$0102 & 2070$\pm$194 & 4069$\pm$255 & 0.15$\pm$0.03 & 6$\pm$1 & 0.9$\pm$0.2 & Y & 1742$\pm$106 & 3145$\pm$145 & 0.10$\pm$0.01 & 11$\pm$0.9 & 1.67$\pm$0.09 & Y \\
0150+0057 & 1799$\pm$172 & 4816$\pm$408 & 0.09$\pm$0.01 & 25$\pm$4 & 2.0$\pm$0.2 & N & 2057$\pm$129 & 4212$\pm$242 & 0.07$\pm$0.01 & 1$\pm$4 & 3.7$\pm$0.3 & Y \\
0206$-$0017 & 2514$\pm$483 & 5054$\pm$761 & 0.12$\pm$0.03 & 18$\pm$5 & 1.68$\pm$0.18 & Y & 1979$\pm$99 & 4060$\pm$148 & 0.10$\pm$0.01 & 34$\pm$3 & 4.7$\pm$0.3 & Y \\
0212+1406 & 1947$\pm$198 & 3776$\pm$299 & 0.11$\pm$0.01 & 15.0$\pm$1.9 & 1.6$\pm$0.1 & Y & 1586$\pm$129 & 2601$\pm$242 & 0.15$\pm$0.01 & 17.7$\pm$1.7 & 2.6$\pm$0.2 & Y\\
0301+0110 & 2078$\pm$226 & 3751$\pm$370 & 0.47$\pm$0.07 & 4.9$\pm$0.8 & 0.90$\pm$0.08 & Y & 1423$\pm$160 & 3612$\pm$442 & 0.42$\pm$0.06 & 6.77$\pm$1.08 & 1.16$\pm$0.11 & Y \\
0301+0115 & 1510$\pm$182 & 3928$\pm$181 & 0.05$\pm$0.01 & 39$\pm$6 & 6.2$\pm$0.5 & Y & 1653$\pm$105 & 3594$\pm$200 & 0.09$\pm$0.01 & 16.2$\pm$1.6 & 2.73$\pm$0.17 & Y\\
0310$-$0049 & 1713$\pm$111 & 3152$\pm$178 & 0.09$\pm$0.01 & 44$\pm$6 & 2.9$\pm$0.4 & Y & 1558$\pm$69 & 2843$\pm$153 & 0.08$\pm$0.02 & 70$\pm$12 & 4.9$\pm$0.8 & Y \\
0336$-$0706 & 3189$\pm$177 & 6827$\pm$416 & 0.19$\pm$0.02 & 4.8$\pm$0.4 & 0.32$\pm$0.02 & Y & 2403$\pm$164 & 7238$\pm$552 & 0.16$\pm$0.04 & 3.6$\pm$1.3 & 0.24$\pm$0.06 & Y \\
0353$-$0623 & 1725$\pm$359 & 4209$\pm$557 & 0.12$\pm$0.03 & 12$\pm$4 & 1.0$\pm$0.2 & Y & 1548$\pm$225 & 3050$\pm$312 & 0.13$\pm$0.01 & 20$\pm$4 & 2.12$\pm$0.15 & Y \\
0731+4522 & 1778$\pm$107 & 3291$\pm$260 & 0.18$\pm$0.01 & 3.3$\pm$0.3 & 0.53$\pm$0.03 & N & 1885$\pm$134 & 3715$\pm$389 & 0.16$\pm$0.01 & 4.5$\pm$0.5 & 0.72$\pm$0.04 & Y \\
0735+3752 & 3120$\pm$303 & 4572$\pm$712 & 0.18$\pm$0.06 & 7$\pm$3 & 0.37$\pm$0.13 & Y & 3996$\pm$293 & 8070$\pm$838 & 0.14$\pm$0.06 & 7$\pm$4 & 0.30$\pm$0.12 & N \\
0737+4244 & 1624$\pm$74 & 4326$\pm$133 & 0.22$\pm$0.01 & 3.8$\pm$0.2 & 0.41$\pm$0.02 & N & 1692$\pm$98 & 3361$\pm$148 & 0.19$\pm$0.01 & 6.2$\pm$0.5 & 0.82$\pm$0.05 & Y \\
0802+3104 & 2661$\pm$265 & 5707$\pm$652 & 0.10$\pm$0.02 & 34$\pm$7 & 2.0$\pm$0.4 & Y & 1772$\pm$185 & 4430$\pm$358 & 0.07$\pm$0.01 & 19.5$\pm$1.5 & 1.66$\pm$0.11 & Y \\
0811+1739 & 1779$\pm$66 & 4325$\pm$156 & 0.06$\pm$0.01 & 41$\pm$6 & 2.4$\pm$0.3 & Y & 1520$\pm$150 & 3520$\pm$253 & 0.11$\pm$0.02 & 31$\pm$9 & 2.6$\pm$0.5 & Y \\
0813+4608 & 1270$\pm$197 & 2992$\pm$471 & 0.08$\pm$0.08 & 5$\pm$7 & 0.3$\pm$0.3 & Y & 1430$\pm$91 & 2483$\pm$165 & 0.11$\pm$0.01 & 14.0$\pm$1.5 & 1.39$\pm$0.11 & Y \\
0831+0521 & 1040$\pm$328 & 1197$\pm$550 & 0.10$\pm$0.02 & 1.3$\pm$0.6 & 0.32$\pm$0.08 & N & \nodata & \nodata & 0.11$\pm$0.06 & [0.06] & [0.32] & N\\
0845+3409 & 2090$\pm$294 & 3897$\pm$865 & 0.16$\pm$0.02 & 9.7$\pm$1.7 & 0.69$\pm$0.07 & Y & 1718$\pm$172 & 2820$\pm$310 & 0.16$\pm$0.02 & 7.5$\pm$1.2 & 0.69$\pm$0.08 & Y \\
0846+2522 & 2572$\pm$244 & 7398$\pm$410 & 0.09$\pm$0.02 & 20$\pm$3 & 1.04$\pm$0.14 & Y & 3394$\pm$402 & 8304$\pm$459 & 0.08$\pm$0.02 & 22$\pm$6 & 0.97$\pm$0.15 & Y \\
0847+1824 & 1665$\pm$193 & 3519$\pm$282 & 0.40$\pm$0.04 & 8.0$\pm$1.1 & 1.19$\pm$0.08 & Y & \nodata & \nodata & 0.27$\pm$0.11 & [1.66] & [0.08] & N \\
0854+1741 & 2729$\pm$353 & 4642$\pm$602 & 0.16$\pm$0.02 & 5.67$\pm$1.08 & 0.84$\pm$0.11 & Y & 1472$\pm$269 & 2582$\pm$563 & 0.22$\pm$0.01 & 5$\pm$1 & 1.36$\pm$0.04 & Y\\
0857+0528 & 1959$\pm$193 & 4307$\pm$282 & 0.17$\pm$0.03 & 22$\pm$4 & 1.8$\pm$0.3 & Y & 1485$\pm$120 & 3499$\pm$51 & 0.1$\pm$0.01 & 17.0$\pm$1.3 & 1.85$\pm$0.13 & Y \\
0904+5536 & 2643$\pm$70 & 5724$\pm$125 & 0.26$\pm$0.02 & 7.8$\pm$0.4 & 0.48$\pm$0.03 & Y & 2483$\pm$36 & 7540$\pm$182 & 0.22$\pm$0.02 & 7.5$\pm$0.5 & 0.45$\pm$0.03 & Y \\
0909+1330 & 1721$\pm$251 & 4023$\pm$416 & 0.12$\pm$0.02 & 18$\pm$4 & 2.00$\pm$0.19 & Y & 1687$\pm$138 & 4418$\pm$148 & 0.19$\pm$0.04 & 45.$\pm$8 & 4.4$\pm$0.5 & Y \\
0921+1017 & 2033$\pm$157 & 4900$\pm$498 & 0.23$\pm$0.04 & 4.6$\pm$0.9 & 0.31$\pm$0.05 & Y & 2317$\pm$286 & 4432$\pm$427 & 0.17$\pm$0.02 & 8.9$\pm$1.4 & 0.71$\pm$0.06 & Y \\
0923+2254 & 2195$\pm$254 & 3783$\pm$545 & 0.51$\pm$0.13 & 4.1$\pm$0.8 & 0.67$\pm$0.15 & Y & 1824$\pm$265 & 2924$\pm$668 & 0.24$\pm$0.03 & 9.4$\pm$2.0 & 1.66$\pm$0.19 & Y \\
0923+2946 & 2686$\pm$222 & 5946$\pm$522 & 0.13$\pm$0.02 & 5.9$\pm$1.1 & 0.24$\pm$0.03 & Y & 2936$\pm$247 & 5650$\pm$906 & 0.16$\pm$0.02 & 11.2$\pm$1.7 & 0.56$\pm$0.06 & Y \\
0927+2301 & 2624$\pm$226 & 7732$\pm$647 & 0.10$\pm$0.03 & 17$\pm$5 & 1.0$\pm$0.2 & Y & 2112$\pm$205 & 5377$\pm$367 & 0.08$\pm$0.01 & 17$\pm$2 & 1.5$\pm$0.1 & Y \\
0932+0233 & 2407$\pm$429 & 6863$\pm$625 & 0.13$\pm$0.01 & 10$\pm$2 &  0.50$\pm$0.04 & Y & 1814$\pm$72 & 4273$\pm$168 & 0.13$\pm$0.01 & 12.54$\pm$1.04 & 1.05$\pm$0.07 & Y \\
0932+0405 & 1829$\pm$75 & 5316$\pm$427 & 0.26$\pm$0.02 & 2.5$\pm$0.2 & 0.24$\pm$0.02 & N & \nodata & \nodata & 0.1$\pm$0.4 &  [0.41] & [0] & N\\
0936+1014 & 2259$\pm$153 & 4846$\pm$228 & 0.09$\pm$0.01 &  9.9$\pm$0.8 & 1.35$\pm$0.07 & Y & 1995$\pm$80 & 3916$\pm$123 &  0.10$\pm$0.01 & 17.02$\pm$1.07 & 2.80$\pm$0.14 & Y\\
0938+0743 & 1663$\pm$190 & 4813$\pm$684 & 0.14$\pm$0.02 & 11$\pm$2 & 1.04$\pm$0.12 & N & 3723$\pm$608 & 7110$\pm$808 &  0.15$\pm$0.12 & 4$\pm$4 & 0.22$\pm$0.17 & N \\
0948+4030 & 1768$\pm$225 & 3460$\pm$258 & 0.17$\pm$0.05 & 12$\pm$4 & 1.2$\pm$0.4 & Y & 3188$\pm$438 & 6732$\pm$548 &  0.12$\pm$0.01 & 11.0$\pm$1.8 &  0.57$\pm$0.05 & Y\\
1002+2648 & 1944$\pm$184 & 5721$\pm$433 & 0.12$\pm$0.02 & 4.6$\pm$1.0 & 0.29$\pm$0.04 & Y & \nodata & \nodata &  0.12$\pm$0.04 & [2.23] & [0.15] & N\\
1029+1408 & 2031$\pm$338 & 5264$\pm$381 & 0.14$\pm$0.02 & 9.8$\pm$2.0 & 0.80$\pm$0.08 & Y & 2456$\pm$344 & 6499$\pm$649 &  0.15$\pm$0.01 & 8.7$\pm$1.5 & 0.59$\pm$0.05 & Y\\
1029+2728 & 2103$\pm$233 & 4958$\pm$547 & 0.23$\pm$0.06 & 4.1$\pm$1.5 & 0.32$\pm$0.08 & Y & 1544$\pm$28 & 3634$\pm$152 &  0.25$\pm$0.03 & 5.07$\pm$1.06 & 0.59$\pm$0.06 & Y\\
1029+4019 & 2515$\pm$349 & 5985$\pm$401 & 0.11$\pm$0.02 & 8.4$\pm$1.5 & 0.64$\pm$0.07 & Y & 2193$\pm$387 & 5998$\pm$547 &  0.12$\pm$0.01 & 6.5$\pm$1.3 & 0.52$\pm$0.04 & Y\\
1038+4658 & 2621$\pm$322 & 4750$\pm$302 & 0.10$\pm$0.01 & 9.4$\pm$1.4 & 0.62$\pm$0.06 & Y & \nodata & \nodata &  0.10$\pm$0.14 & [2.75] & [0.14] & N\\
1042+0414 & 1569$\pm$194 & 4064$\pm$166 & 0.34$\pm$0.03 & 6.6$\pm$1.0 & 0.57$\pm$0.04 & Y & 1518$\pm$102 & 2724$\pm$145 & 0.33$\pm$0.03 & 5.6$\pm$0.6 & 0.73$\pm$0.05 & Y\\
1043+1105 & 2864$\pm$149 & 6171$\pm$230 & 0.21$\pm$0.04 & 10.5$\pm$1.7 & 0.61$\pm$0.06 & Y & 2313$\pm$28 & 6597$\pm$152 & 0.13$\pm$0.01 & 4.7$\pm$0.2 & 0.25$\pm$0.02 & Y\\
1049+2451 & 2368$\pm$161 & 5181$\pm$207 & 0.15$\pm$0.03 & 14.2$\pm$1.8 & 0.97$\pm$0.14 & Y & 2534$\pm$135 & 5112$\pm$203 & 0.15$\pm$0.01 & 15.1$\pm$1.2 & 1.04$\pm$0.07 & Y\\
1058+5259 & 1987$\pm$400 & 4928$\pm$940 & 0.07$\pm$0.01 & 18$\pm$5 & 1.24$\pm$0.18 & N & 1896$\pm$150 & 5372$\pm$150 & 0.10$\pm$0.01 & 13.3$\pm$1.7 & 0.92$\pm$0.08 & Y\\
1101+1102 & 2558$\pm$125 & 6047$\pm$295 & 0.14$\pm$0.03 & 7.6$\pm$1.2 & 0.62$\pm$0.13 & N & 3949$\pm$170 & 8349$\pm$597 & 0.13$\pm$0.01 & 3.5$\pm$0.3 & 0.20$\pm$0.01 & Y\\
1104+4334 & 1873$\pm$308 & 4319$\pm$647 & 0.09$\pm$0.02 & 4.4$\pm$1.3 & 0.31$\pm$0.06 & Y & 1719$\pm$160 & 4072$\pm$395 & 0.12$\pm$0.01 & 9.4$\pm$1.2 & 0.84$\pm$0.07 & Y\\
1110+1136 & 1878$\pm$206 & 3860$\pm$400 & 0.14$\pm$0.01 & 10.4$\pm$1.6 & 0.78$\pm$0.05 & Y & 1378$\pm$95 & 2898$\pm$150 & 0.15$\pm$0.01 & 14.7$\pm$1.5 & 1.68$\pm$0.12 & Y\\
1116+4123 & 2531$\pm$294 & 6324$\pm$692 & 0.25$\pm$0.04 & 10.2$\pm$1.9 & 0.56$\pm$0.07 & N & 3136$\pm$315 & 6774$\pm$740 & 0.27$\pm$0.04 & 6.6$\pm$1.3 & 0.39$\pm$0.05 & Y\\
1118+2827 & 1908$\pm$136 & 5910$\pm$498 & 0.21$\pm$0.03 & 3.8$\pm$0.7 & 0.38$\pm$0.05 & N & \nodata & \nodata & 0.14$\pm$0.05 & [2.05] & [0.19] & N\\
1132+1017 & 2028$\pm$147 & 5782$\pm$345 & 0.10$\pm$0.01 & 7.7$\pm$1.3 & 0.74$\pm$0.09 & Y & 1900$\pm$86 & 5341$\pm$740 & 0.09$\pm$0.01 & 15.8$\pm$1.4 & 1.63$\pm$0.11 & Y\\
1137+4826 & 1750$\pm$357 & 3663$\pm$647 & 0.34$\pm$0.05 & 5.8$\pm$1.4 & 0.94$\pm$0.14 & Y & 1606$\pm$92 & 3788$\pm$222 & 0.45$\pm$0.06 & 6.8$\pm$0.9 & 1.00$\pm$0.09 & Y\\
1139+5911 & 2333$\pm$158 & 4262$\pm$218 & 0.16$\pm$0.06 & 20$\pm$4 & 2.2$\pm$0.7 & Y & 2228$\pm$111 & 3994$\pm$221 & 0.09$\pm$0.01 & 22.2$\pm$1.7 & 1.73$\pm$0.11 & Y\\
1140+2307 & 2710$\pm$235 & 5014$\pm$668 & 0.10$\pm$0.01 & 8.0$\pm$1.2 & 0.43$\pm$0.05 & N & 3324$\pm$330 & 6376$\pm$586 & 0.12$\pm$0.01 & 4.4$\pm$0.6 & 0.24$\pm$0.02 & N\\
1143+5941 & 2002$\pm$446 & 5629$\pm$863 & 0.13$\pm$0.04 & 15.0$\pm$2.7 & 0.77$\pm$0.18 & Y & 1790$\pm$128 & 5405$\pm$424 & 0.08$\pm$0.01 & 38$\pm$4 & 2.00$\pm$0.08 & Y\\
1144+3653 & 3016$\pm$292 & 8301$\pm$579 & 0.08$\pm$0.01 & 21$\pm$3 & 0.90$\pm$0.14 & N & 2933$\pm$205 & 8009$\pm$145 & 0.08$\pm$0.01 & 16.3$\pm$1.5 & 0.87$\pm$0.05 & N\\
1145+5547 & 2078$\pm$422 & 4298$\pm$611 & 0.10$\pm$0.01 & 14$\pm$3 & 1.18$\pm$0.12 & Y & 1837$\pm$208 & 4465$\pm$391 & 0.15$\pm$0.01 & 7.49$\pm$1.06 & 0.71$\pm$0.05 & Y\\
1147+0902 & 3733$\pm$226 & 6475$\pm$563 & 0.14$\pm$0.05 & 13$\pm$3 & 0.9$\pm$0.3 & N & 2896$\pm$188 & 5285$\pm$543 & 0.12$\pm$0.01 & 10.9$\pm$1.2 & 0.73$\pm$0.06 & Y\\
1205+4959 & 3572$\pm$201 & 8275$\pm$650 & 0.11$\pm$0.02 & 3.9$\pm$0.4 & 0.24$\pm$0.03 & Y & 2678$\pm$294 & 5552$\pm$374 & 0.10$\pm$0.01 & 3.6$\pm$0.5 & 0.31$\pm$0.02 & Y\\
1206+4244 & 1925$\pm$167 & 3889$\pm$211 & 0.15$\pm$0.02 & 24$\pm$3 & 2.4$\pm$0.3 & Y & 1614$\pm$92 & 3800$\pm$144 & 0.17$\pm$0.02 & 36$\pm$3 & 3.6$\pm$0.3 & Y\\
1210+3820 & 2499$\pm$432 & 6413$\pm$513 & 0.20$\pm$0.04 & 5.9$\pm$1.5 & 0.36$\pm$0.06 & N & 2831$\pm$148 & 5300$\pm$392 & 0.20$\pm$0.02 & 8.0$\pm$0.7 & 0.58$\pm$0.04 & Y\\
1216+5049 & 3329$\pm$180 & 8923$\pm$422 & 0.11$\pm$0.02 & 2.3$\pm$0.3 & 0.19$\pm$0.03 & Y & 4487$\pm$477 & 7810$\pm$392 & 0.09$\pm$0.01 & 2.1$\pm$0.3 & 0.15$\pm$0.01 & Y\\
1223+0240 & 2780$\pm$160 & 5802$\pm$220 & 0.05$\pm$0.02 & 207$\pm$44 & 17$\pm$6 & Y & 2306$\pm$107 & 5051$\pm$168 & 0.10$\pm$0.01 & 69$\pm$6 & 5.7$\pm$0.4 & Y\\
1228+0951 & 2289$\pm$657 & 7303$\pm$1543 & 0.12$\pm$0.07 & 3$\pm$3 & 0.23$\pm$0.14 & Y & 2331$\pm$456 & 6011$\pm$495 & 0.12$\pm$0.04 & 3.2$\pm$1.6 & 0.28$\pm$0.10 & Y\\
1231+4504 & 1551$\pm$343 & 2872$\pm$440 & 0.20$\pm$0.03 & 5.5$\pm$1.4 & 1.27$\pm$0.14 & Y & 1555$\pm$168 & 3325$\pm$394 & 0.16$\pm$0.02 & 9.1$\pm$1.2 & 1.91$\pm$0.16 & Y\\
1241+3722 & 1829$\pm$93 & 4320$\pm$219 & 0.13$\pm$0.02 &  7.9$\pm$0.9 & 0.69$\pm$0.08 & N & 1574$\pm$100 & 3185$\pm$197 & 0.11$\pm$0.01 & 6.0$\pm$0.8 & 0.70$\pm$0.07 & Y\\
1246+5134 & 2402$\pm$313 & 4403$\pm$703 & 0.07$\pm$0.05 & 10$\pm$11 & 0.6$\pm$0.5 & Y & 1141$\pm$130 & 2270$\pm$185 & 0.09$\pm$0.01 & 12.0$\pm$1.8 & 1.52$\pm$0.12 & Y\\
1250$-$0249 & 2068$\pm$323 & 5304$\pm$732 & 0.17$\pm$0.03 & 9$\pm$2 & 0.89$\pm$0.13 & Y & 2417$\pm$246 & 5771$\pm$541 & 0.18$\pm$0.01 & 7.3$\pm$0.9 & 0.61$\pm$0.04 & Y\\
1306+4552 & 1327$\pm$148 & 3262$\pm$237 & 0.15$\pm$0.02 & 18$\pm$3 & 1.9$\pm$0.2 & N & 1892$\pm$297 & 4129$\pm$772 & 0.22$\pm$0.04 & 2.8$\pm$0.8 & 0.27$\pm$0.05 & Y\\
1307+0952 &  1616$\pm$114 & 3748$\pm$267 & 0.09$\pm$0.04 & 11$\pm$7 & 0.8$\pm$0.3 & N & 1630$\pm$165 & 3586$\pm$249 & 0.14$\pm$0.01 & 12.7$\pm$1.7 & 1.10$\pm$0.09 & N\\
1312+2628 & 1585$\pm$171 & 3131$\pm$256 & 0.12$\pm$0.03 & 32$\pm$10 & 2.8$\pm$0.5 & Y & 1572$\pm$150 & 2924$\pm$345 & 0.22$\pm$0.03 & 36$\pm$6 & 2.95$\pm$0.17 & Y\\
1313+3653 & 2592$\pm$217 & 5591$\pm$402 & 0.14$\pm$0.02 & 4.8$\pm$0.5 & 0.34$\pm$0.03 & Y & 2115$\pm$264 & 4920$\pm$347 & 0.13$\pm$0.01 & 4.2$\pm$0.6 & 0.34$\pm$0.02 & Y\\
1323+2701 & 4266$\pm$349 & 10123$\pm$299 & 0.089$\pm$0.004 & 18.3$\pm$1.7 & 0.79$\pm$0.02 & Y & 2414$\pm$376 & 5219$\pm$782 & 0.10$\pm$0.01 & 5.9$\pm$1.2 & 0.54$\pm$0.06 & Y\\
1353+3951 & 2037$\pm$121 & 6308$\pm$537 & 0.19$\pm$0.02 & 4.1$\pm$0.5 & 0.31$\pm$0.02 & N & \nodata & \nodata & 0.2$\pm$0.7 & [1.52] & [0.31] & N\\
1355+3834 & 2842$\pm$79 & 5936$\pm$245 & 0.28$\pm$0.05 & 3.3$\pm$0.3 & 0.36$\pm$0.06 & Y & 4034$\pm$301 & 6371$\pm$277 & 0.20$\pm$0.02 & 3.2$\pm$0.3 & 0.22$\pm$0.01 & N\\
1405$-$0259 & 1873$\pm$212 & 3518$\pm$549 & 0.21$\pm$0.02 & 13$\pm$2 & 1.48$\pm$0.13 & Y & 1599$\pm$140 & 2933$\pm$260 & 0.16$\pm$0.02 & 11.1$\pm$1.4 & 1.48$\pm$0.12 & Y\\
1416+0317 & 11853$\pm$251 & 4022$\pm$589 & 0.12$\pm$0.02 & 5.4$\pm$1.1 & 0.78$\pm$0.09 & Y & 1514$\pm$233 & 3565$\pm$550 & 0.11$\pm$0.02 & 2.2$\pm$0.5 & 0.40$\pm$0.05 & Y\\
1419+0754 & 1940$\pm$154 & 5517$\pm$362 & 0.10$\pm$0.02 & 2.9$\pm$0.7 & 0.35$\pm$0.06 & N & 3006$\pm$371 & 5780$\pm$529 & 0.10$\pm$0.02 & 3.5$\pm$1.0 & 0.36$\pm$0.07 & Y\\
1423+2720 & 3428$\pm$264 & 7873$\pm$621 & 0.11$\pm$0.03 &  12$\pm$4 & 0.5$\pm$0.1 & Y & \nodata & \nodata & 0.08$\pm$0.04 & [2.45] & [0] & Y\\
1434+4839 & 2268$\pm$171 & 4351$\pm$534 & 0.08$\pm$0.02 & 35$\pm$7 & 2.7$\pm$0.6 & Y & 1731$\pm$85 & 4475$\pm$222 & 0.09$\pm$0.01 & 6.0$\pm$0.4 & 0.58$\pm$0.03 & Y\\
1505+0342 & 2280$\pm$215 & 5028$\pm$569 & 0.14$\pm$0.03 & 6.9$\pm$1.3 & 0.83$\pm$0.12 & N & 1956$\pm$139 & 5782$\pm$154 & 0.08$\pm$0.01 & 17.8$\pm$1.6 & 1.55$\pm$0.09 & Y\\
1535+5754 & 2431$\pm$311 & 4191$\pm$565 & 0.09$\pm$0.03 & 16$\pm$5 & 1.4$\pm$0.4 & Y & 2442$\pm$93 & 5088$\pm$127 & 0.08$\pm$0.01 & 16.2$\pm$1.3 & 1.41$\pm$0.07 & Y\\
1543+3631 & 1527$\pm$171 & 2849$\pm$197 & 0.14$\pm$0.02 & 7.1$\pm$1.1 & 1.03$\pm$0.15 & Y & 1820$\pm$168 & 3831$\pm$248 & 0.09$\pm$0.01 & 4.0$\pm$0.5 & 0.47$\pm$0.03 & Y\\
1545+1709 & 2158$\pm$156 & 5555$\pm$275 & 0.09$\pm$0.01 & 6.5$\pm$0.6 & 0.52$\pm$0.05 & N & 3588$\pm$226 & 4612$\pm$237 & 0.06$\pm$0.01 & 24$\pm$3 & 1.68$\pm$0.14 & N\\
1554+3238 & 2067$\pm$104 & 4887$\pm$246 & 0.14$\pm$0.02 & 7.2$\pm$0.8 & 0.83$\pm$0.11 & N & 2523$\pm$159 & 4148$\pm$258 & 0.11$\pm$0.01 & 7.9$\pm$0.7 & 1.04$\pm$0.06 & Y\\
1557+0830 & 13174$\pm$214 & 5054$\pm$185 & 0.18$\pm$0.03 & 11.9$\pm$1.5 & 0.87$\pm$0.09 & Y & 2388$\pm$91 & 4817$\pm$156 & 0.17$\pm$0.02 & 15.1$\pm$1.4 & 1.09$\pm$0.08 & Y\\
1605+3305 & 2153$\pm$101 & 5079$\pm$230 & 0.06$\pm$0.02 & 54$\pm$13 & 2.9$\pm$1.0 & Y & 1960$\pm$272 & 5302$\pm$637 & 0.10$\pm$0.01 & 33$\pm$5 & 2.03$\pm$0.14 & Y\\
1606+3324 & 2158$\pm$170 & 5087$\pm$393 & 0.10$\pm$0.01 & 3.8$\pm$0.5 & 0.34$\pm$0.04 & N & 2053$\pm$80 & 5088$\pm$739 & 0.12$\pm$0.01 & 4.4$\pm$0.3 & 0.45$\pm$0.03 & Y\\
1611+5211 & 1392$\pm$207 & 3895$\pm$486 & 0.16$\pm$0.03 & 4.8$\pm$1.1 & 0.71$\pm$0.10 & N & 2515$\pm$410 & 7695$\pm$964 & 0.12$\pm$0.03 & 2.9$\pm$1.1 & 0.27$\pm$0.07 & Y\\
1636+4202 & 2367$\pm$223 & 6655$\pm$621 & 0.07$\pm$0.01 & 39$\pm$7 & 3.7$\pm$0.4 & Y & 2492$\pm$230 & 4542$\pm$523 & 0.10$\pm$0.01 & 24$\pm$3 & 2.45$\pm$0.19 & Y\\
1647+4442 & 2227$\pm$279 & 6228$\pm$655 & 0.09$\pm$0.02 & 4.8$\pm$1.9 & 1.3$\pm$0.4 & N & 2921$\pm$246 & 8325$\pm$214 & 0.20$\pm$0.03 & 21$\pm$4 & 0.77$\pm$0.08 & N\\
1655+2014 & \nodata & \nodata & \nodata & \nodata & \nodata & N & \nodata & \nodata & \nodata & \nodata & \nodata & N\\
1708+2153 & 2829$\pm$134 & 6055$\pm$580 & 0.11$\pm$0.01 & 15.2$\pm$1.2 & 1.49$\pm$0.08 & Y & 2402$\pm$122 & 7359$\pm$245 & 0.13$\pm$0.02 & 62$\pm$10 & 6.2$\pm$0.6 & Y\\
2116+1102 & 2790$\pm$27 & 6577$\pm$64 & 0.0903$\pm$0.0002 & 2.25$\pm$0.02 & 0.1831$\pm$0.0003 & Y & 2484$\pm$42 & 7186$\pm$248 & 0.084$\pm$0.005 & 3.63$\pm$0.18 & 0.33$\pm$0.01 & Y\\
2140+0025 & 1329$\pm$104 & 2225$\pm$127 & 0.53$\pm$0.09 & 7.1$\pm$1.1 & 1.47$\pm$0.12 & Y & 1114$\pm$64 & 2155$\pm$127 & 0.29$\pm$0.04 & 17$\pm$2 & 3.7$\pm$0.4 & Y\\
2215$-$0036 & 1877$\pm$200 & 3330$\pm$155 & 0.12$\pm$0.01 & 8.5$\pm$1.1 & 1.58$\pm$0.14 & Y & 1636$\pm$92 & 3966$\pm$220 & 0.09$\pm$0.01 & 7.3$\pm$0.6 & 1.14$\pm$0.06 & Y\\
2221$-$0906 & 2498$\pm$394 & 6684$\pm$539 & 0.10$\pm$0.01 & 21$\pm$4 & 1.20$\pm$0.10 & N & 2375$\pm$131 & 6012$\pm$224 & 0.12$\pm$0.01 & 24$\pm$2 & 1.45$\pm$0.11 & Y\\
2222$-$0819 & 1811$\pm$88 & 3327$\pm$319 & 0.25$\pm$0.02 & 2.5$\pm$0.2 & 0.55$\pm$0.03 & N & 1799$\pm$168 & 2861$\pm$343 & 0.2$\pm$0.02 & 2.5$\pm$0.3 & 0.66$\pm$0.03 & Y\\
2233+1312 & 1897$\pm$66 & 4409$\pm$236 & 0.19$\pm$0.02 & 6.7$\pm$0.4 & 0.90$\pm$0.06 & N & 2477$\pm$135 & 5830$\pm$318 & 0.16$\pm$0.01 & 5.7$\pm$0.5 & 0.53$\pm$0.04 & N\\
2254+0046 & 1466$\pm$200 & 2015$\pm$195 & 0.56$\pm$0.09 & 3.9$\pm$0.8 & 1.26$\pm$0.17 & Y & 859$\pm$194 & 1398$\pm$286 & 0.59$\pm$0.07 & 4.7$\pm$1.1 & 1.66$\pm$0.18 & Y\\
2327+1524 & 3267$\pm$206 & 4098$\pm$707 & 0.06$\pm$0.01 & 11.0$\pm$1.6 & 1.10$\pm$0.13 & N & 1924$\pm$166 & 5807$\pm$390 & 0.08$\pm$0.02 & 2.0$\pm$0.7 & 0.22$\pm$0.06 & N\\
2351+1552 & 3533$\pm$269 & 10437$\pm$526 & 0.12$\pm$0.01 & 3.6$\pm$0.4 & 0.20$\pm$0.02 & N & 2974$\pm$144 & 7803$\pm$394 & 0.11$\pm$0.01 & 6.7$\pm$0.6 & 0.43$\pm$0.03 & N
\enddata
\tablecomments{Col. (1): Target ID based on R.A. and Dec. used throughout the text.
  Col. (2): Second moment of broad H$\beta$ from SDSS spectrum (in km\,s$^{-1}$).
  Col. (3): Full-width at half-maximum of broad H$\beta$ from SDSS spectrum (in km\,s$^{-1}$).
  Col. (4): Integrated flux ratio of H$\beta$/[OIII] from SDSS spectrum.
  Col. (5): Integrated flux ratio of H$\beta$ broad/narrow from SDSS spectrum.
  Col. (6): Peak flux ratio of H$\beta$ broad/narrow from SDSS spectrum.
  Col. (7): Whether or not FeII was subtracted from the SDSS spectrum during the fitting process.
  Col. (8): Second moment of broad H$\beta$ from Keck spectrum (in km\,s$^{-1}$).
  Col. (9): Full-width at half-maximum of broad H$\beta$ from Keck spectrum  (in km\,s$^{-1}$).
  Col. (10): Integrated flux ratio of H$\beta$/[OIII] from Keck spectrum.
  Col. (11): Integrated flux ratio of H$\beta$ broad/narrow from Keck spectrum.
  Col. (12): Peak flux ratio of H$\beta$ broad/narrow from Keck spectrum.
  Col. (13):  Whether or not FeII was subtracted from the  Keck spectrum during the fitting process.\\
  Note that values in brackets are estimated from upper limits on the broad H$\beta$ flux for objects classified as Sy 1.9 or Sy 2, that is without an obvious broad H$\beta$ component.
No width (FWHM or $\sigma$) of broad H$\beta$ is given in those cases.
Also note that for object 1655+2014 the signal-to-noise ratio was too low in both SDSS and Keck spectra to produce a reliable fit to the H$\beta$ region. No values are given for this object and it is not included in the quantitative analysis.}
\label{table:fitresults}
\end{deluxetable*}

\section{SDSS vs. Keck Spectra Comparison}
\label{appendix:agnspectra}
\begin{figure}[h!]
\centering
\includegraphics[width=\textwidth]{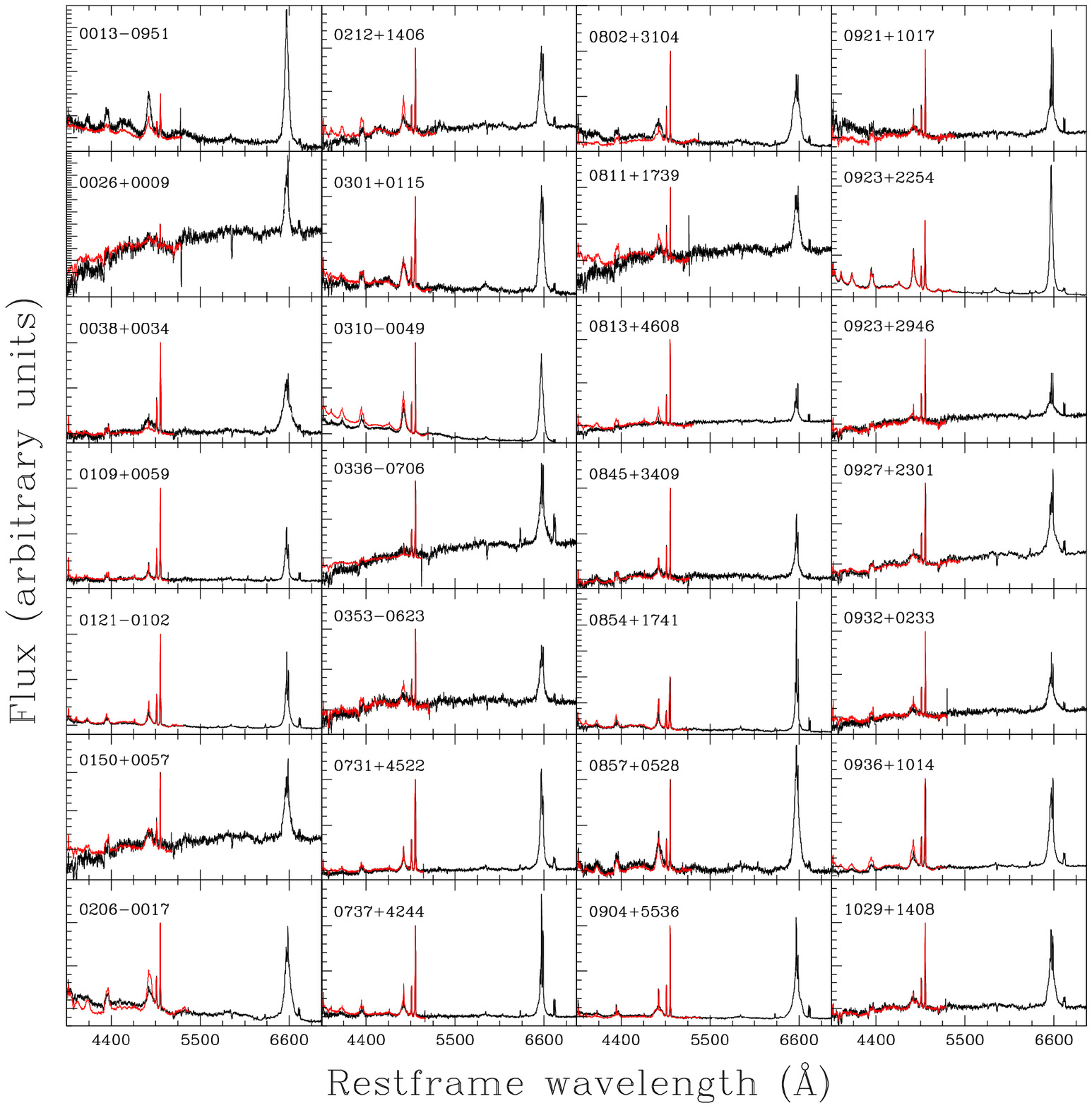}
\caption{Unsubtracted spectra comparing SDSS (black) and Keck (red)
  (3850-7000\AA). Note that the Keck spectra only cover the blue part.
  For comparison, the base of the 5007\AA \space [OIII] line is set to 0 and the peak to 1 to make the data sets comparable as discussed in the text. Objects shown in this figure
  are included in \citet{ben15}.}
\label{fig:hbeta20}
\end{figure}

\begin{figure}
\centering
\includegraphics[width=\textwidth]{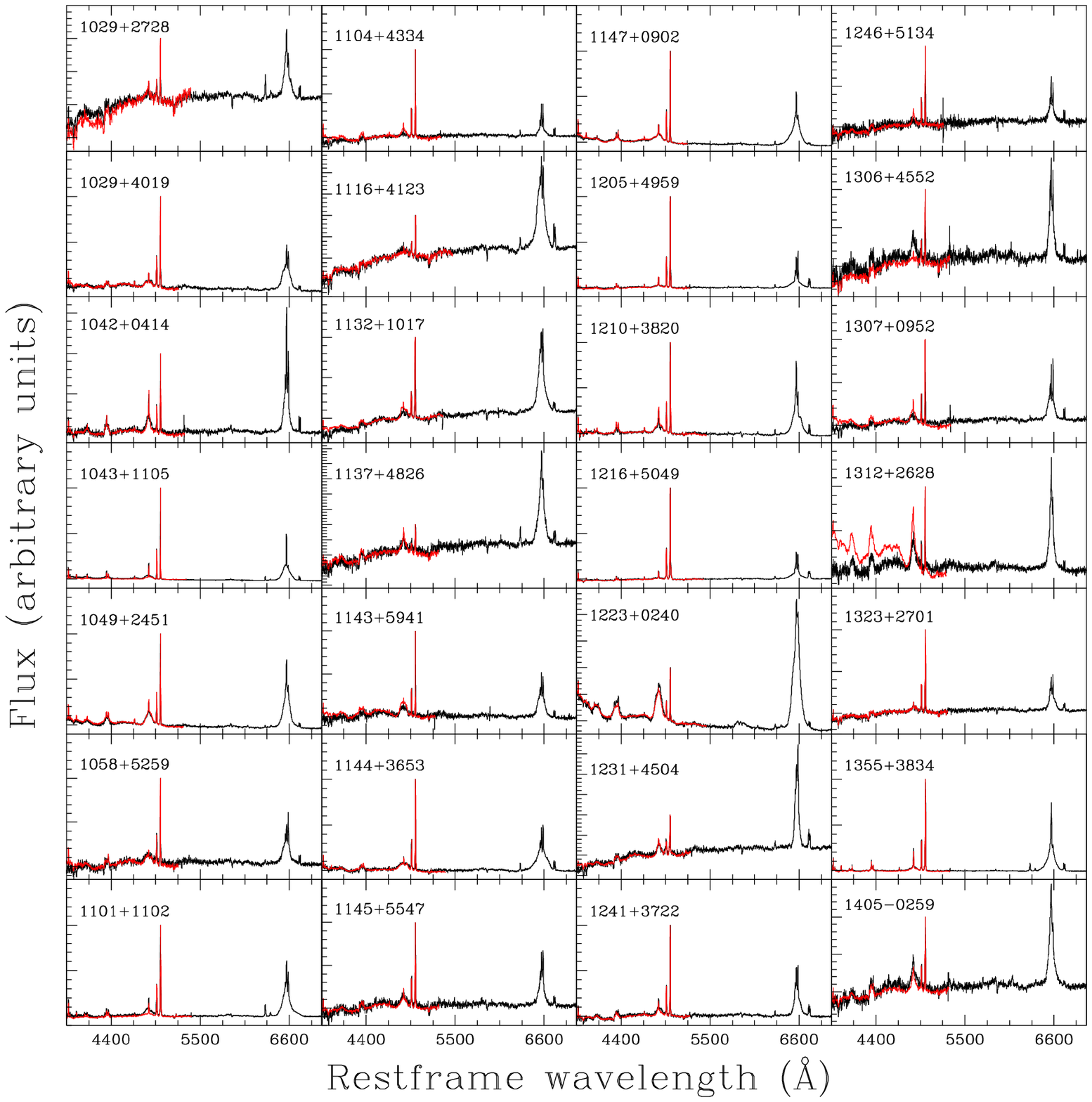}
\caption{Same as Figure~\ref{fig:hbeta20}.}
\label{fig:hbeta21}
\end{figure}

\begin{figure}
\centering
\includegraphics[width=\textwidth]{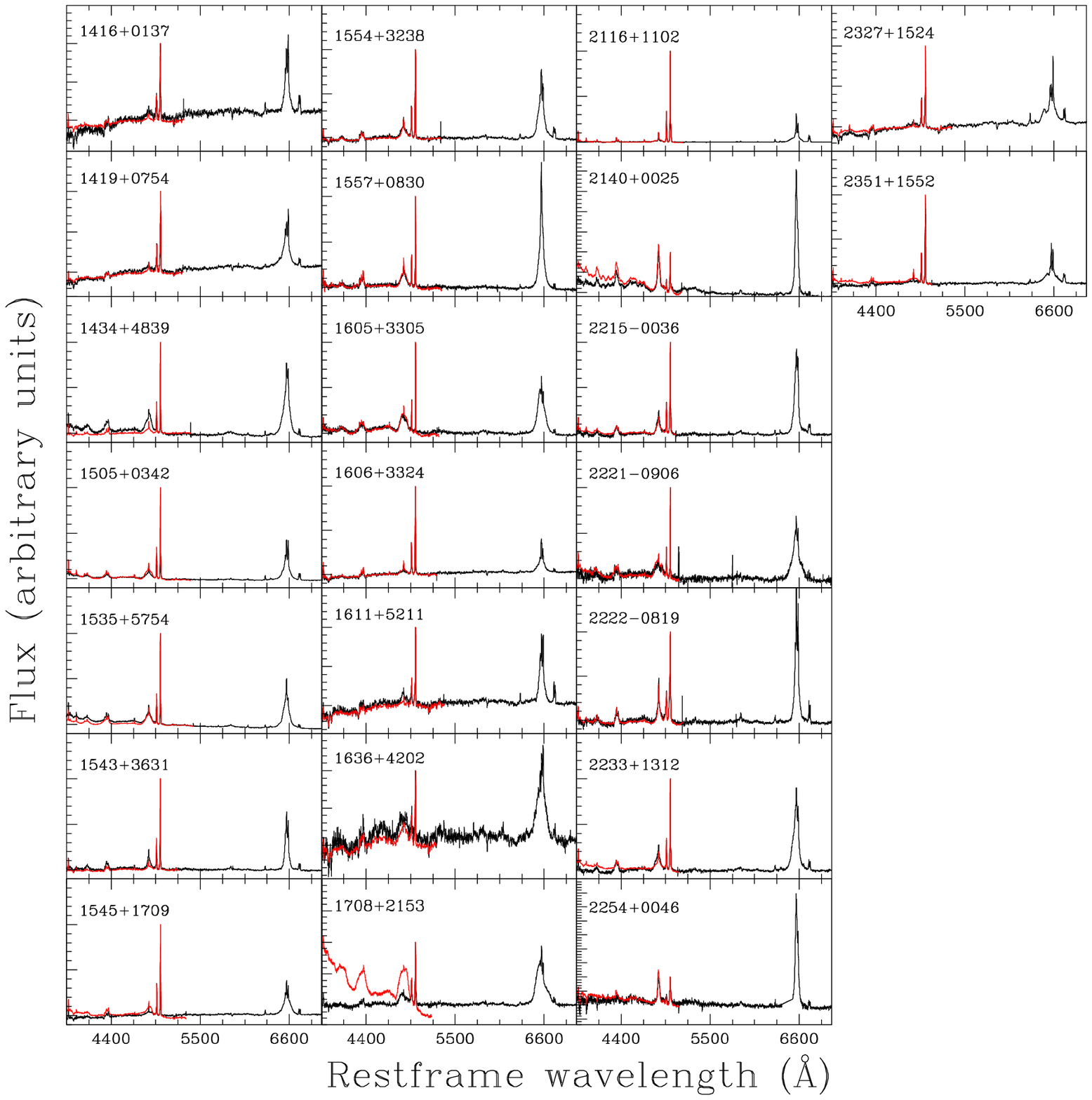}
\caption{Same as Figure~\ref{fig:hbeta20}.}
\label{fig:hbeta22}
\end{figure}

\begin{figure}
\centering
\includegraphics[width=\textwidth]{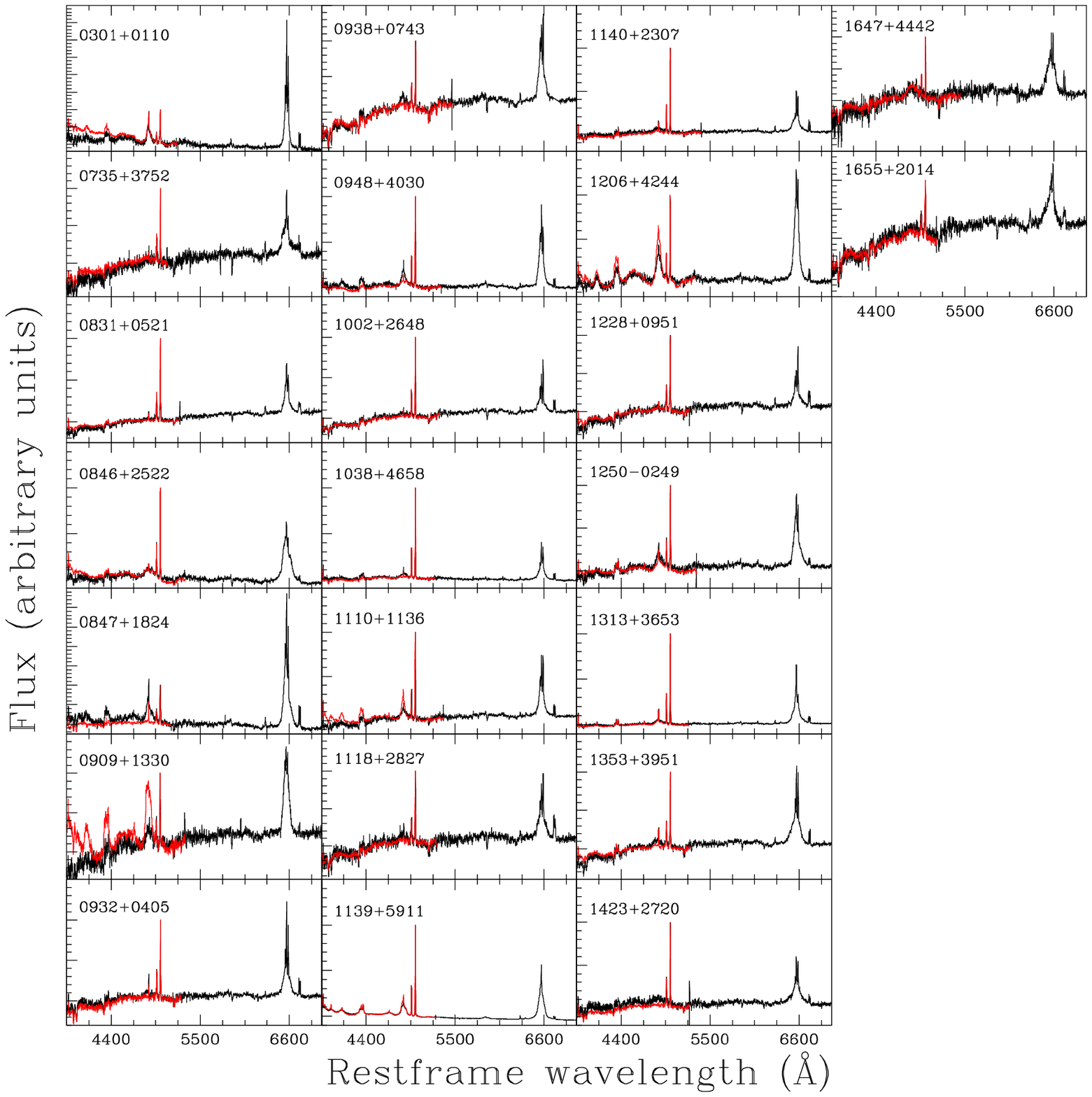}
\caption{Same as Figure~\ref{fig:hbeta20}, except that these are the 23 objects not included in \citet{ben15}.}
\label{fig:hbeta23}
\end{figure}

\section{Fits to SDSS and Keck Spectra}
\label{appendix:fits}
\begin{figure}[h!]
\centering
\includegraphics[scale=0.45]{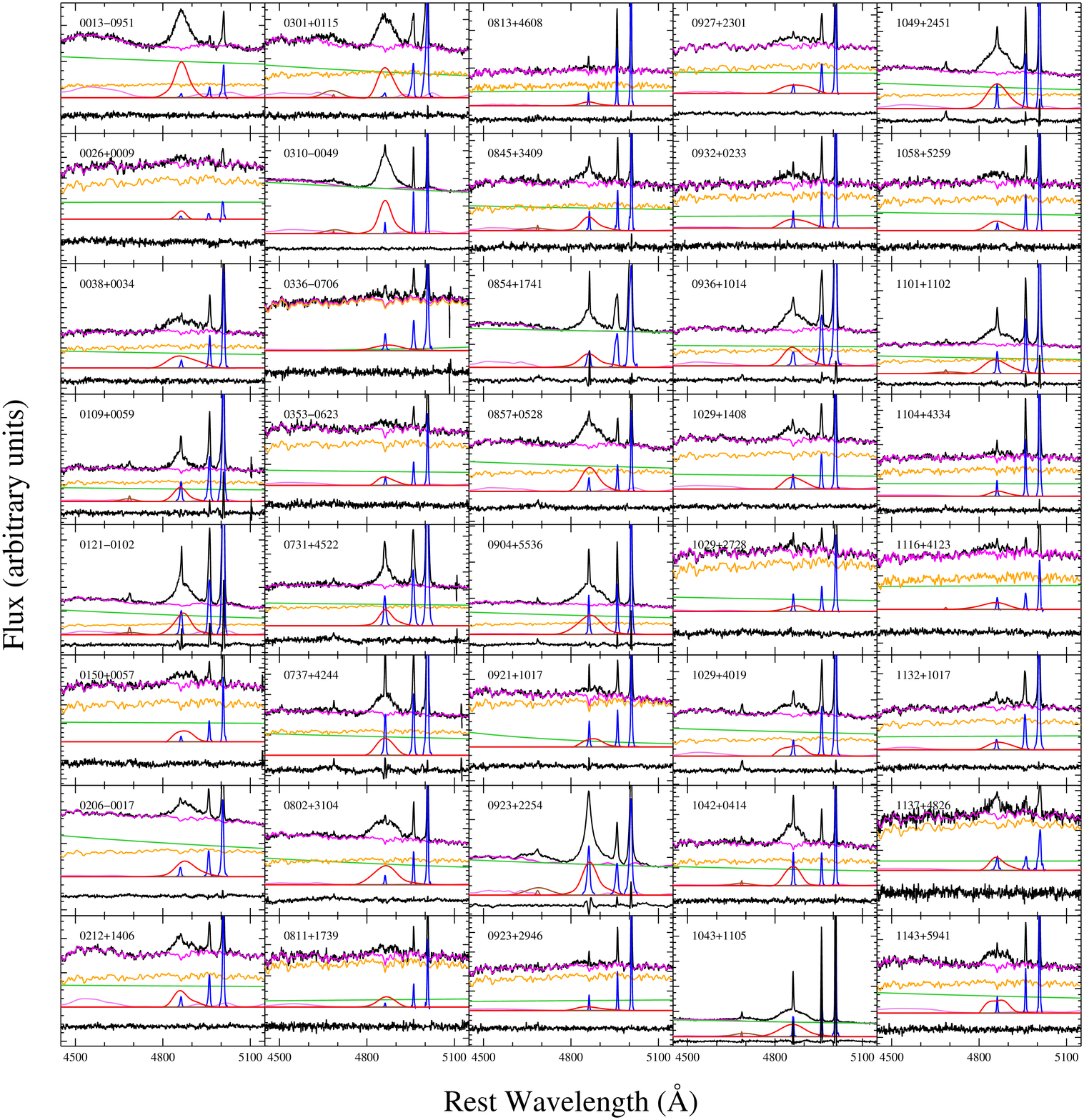}
\caption{Multi-component spectral decomposition
  of the H$\beta$ region in SDSS spectra.
  In the upper region, the observed spectrum is shown in black with the best-fit of the continuum, FeII, and host galaxy starlight in magenta.  Below that the best-fit to the power-law continuum is shown in green with the stellar spectrum in yellow.  Below this, the narrow lines of H$\beta \lambda$4861, and [OIII] $\lambda\lambda$4959,5007 are shown in blue, the broad and narrow components of HeII $\lambda$4686 in brown, and the broad component of H$\beta$ in red. The residuals are plotted in black (arbitrarily shifted downward for clarity).
  Note that the objects shown here are included in \citet{ben15}
  who show the corresponding fit to the Keck spectra in their Figure 3.}
\label{fig:SDSS1}
\end{figure}

\begin{figure}
\centering
\includegraphics[scale=0.45]{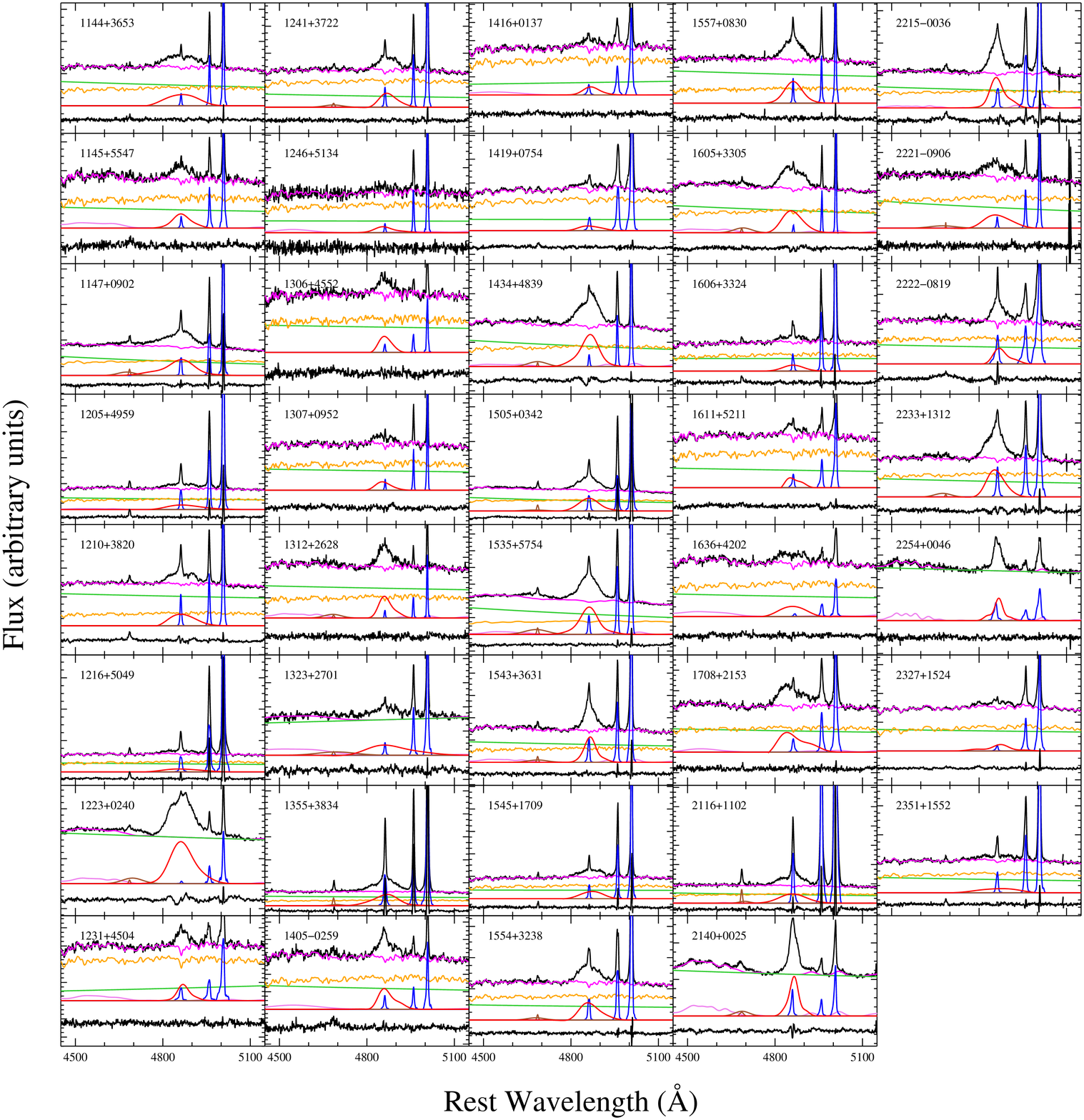}
\caption{Same as Figure~\ref{fig:SDSS1}.}
\label{fig:SDSS2}
\end{figure}

\begin{figure}
\centering
\includegraphics[scale=0.45]{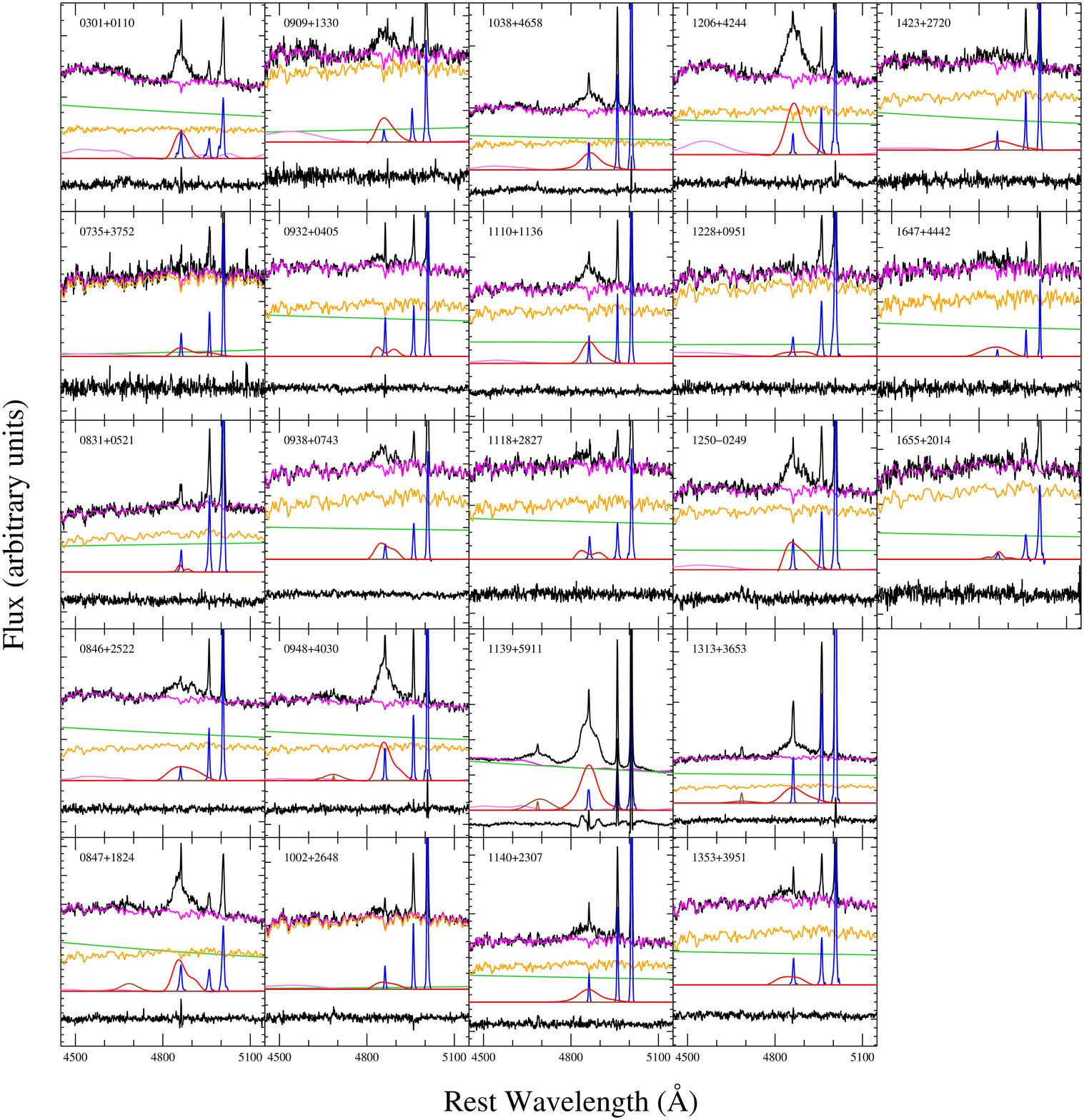}
\caption{Same as Figure~\ref{fig:SDSS1}, but for 23 objects not included in \citet{ben15}.}
\label{fig:SDSS3}
\end{figure}

\begin{figure}
\centering
\includegraphics[scale=0.45]{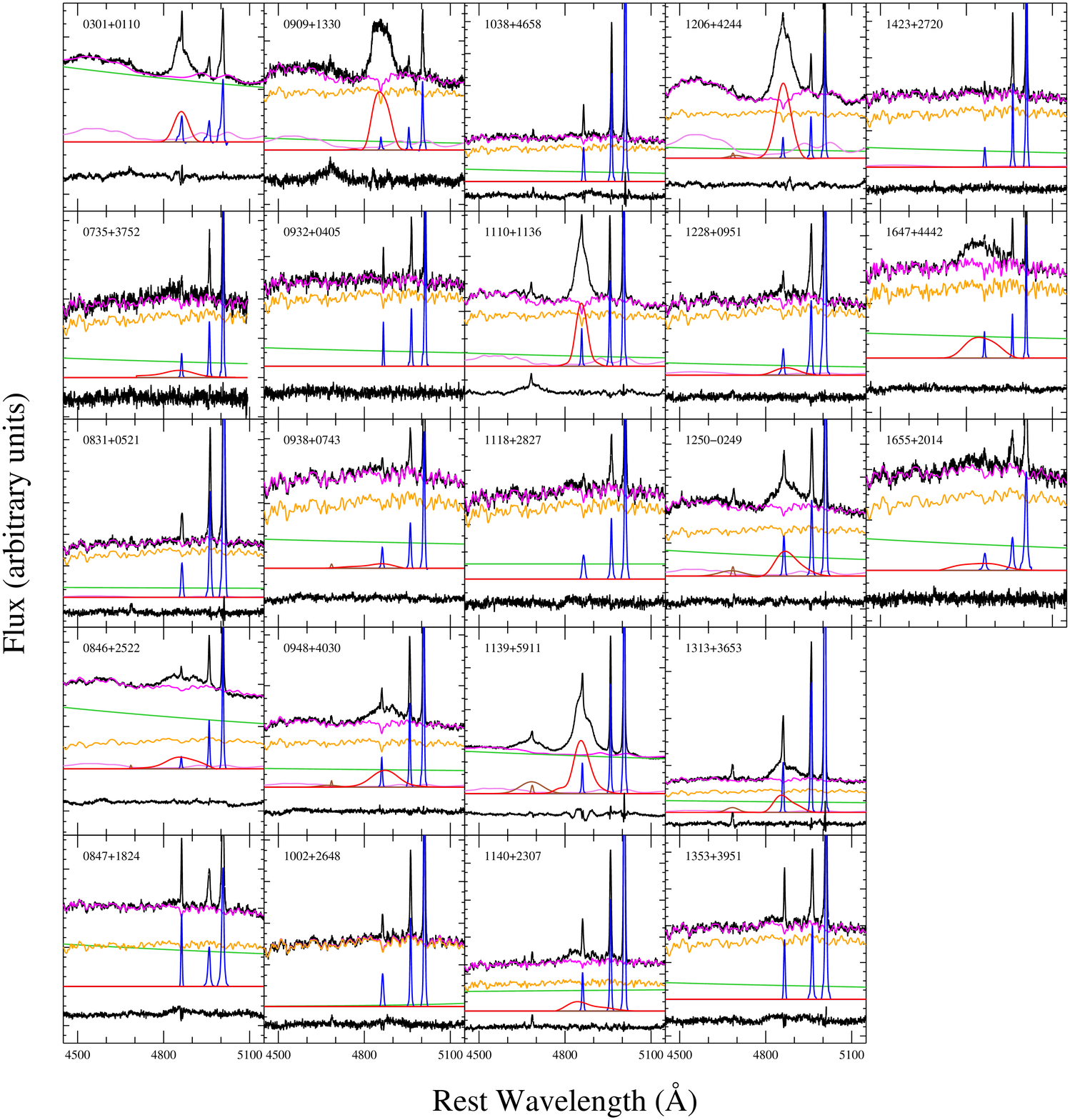}
\caption{Same as Figure~\ref{fig:SDSS3} for spectra gathered from Keck and not included in \cite{ben15}.}
\label{fig:Keck3}
\end{figure}

\section{SDSS Multi-Color Images for 0847+1824 \& 1038+4658}
\label{appendix:sdss}
\begin{figure}[h!]
\centering
\includegraphics[scale=0.6]{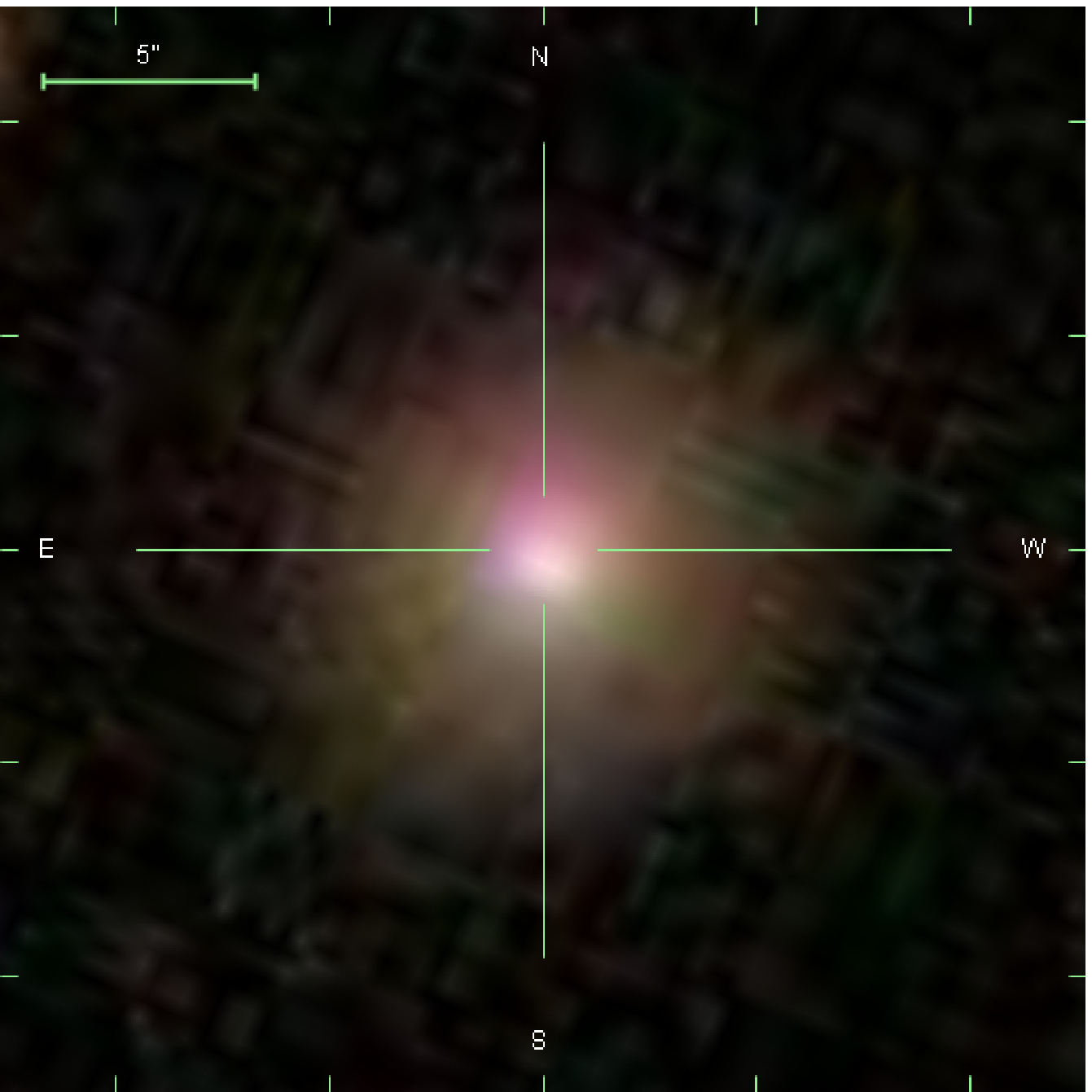}
\includegraphics[scale=0.6]{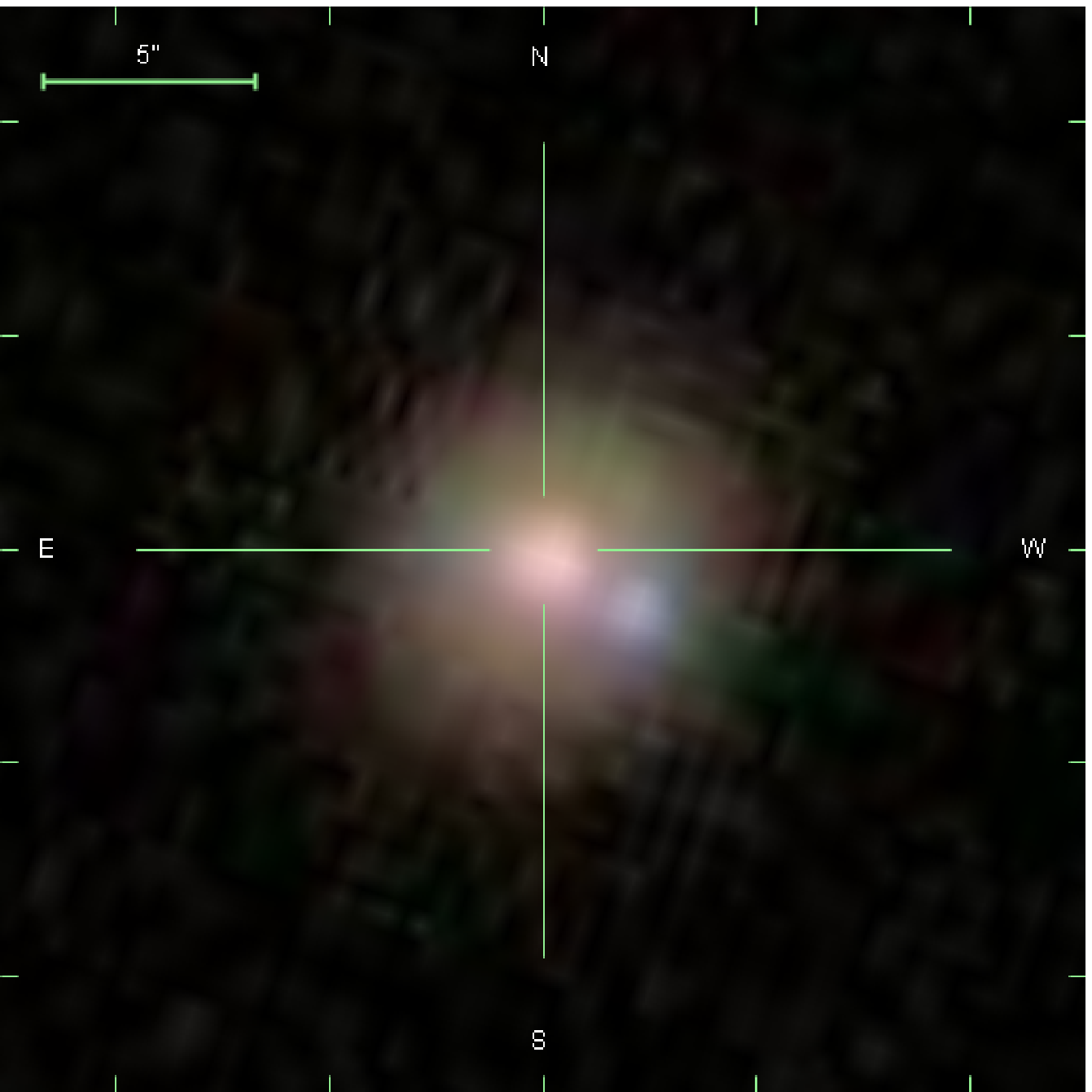}
\caption{SDSS multi-color image for 0847+1824 (left;  taken roughly a year before the SDSS spectrum,
    on 12-13-2004) and 1038+4658 (right; taken roughly 10 months before the SDSS spectrum,
    on 02-08-2002).
  For 0847+1824, there is extended emission offset $\sim$1\arcsec~to the north-east of the galaxy center.
  For 1038+4658, an emission blob can be seen $\sim$2.5\arcsec~to the south-west of the galaxy center.
  In both cases, the extended emission might have been missed in the Keck spectra due to the smaller
  width longslit (with a position angle not covering the emission), but included in the 3\arcsec~fiber of SDSS.
}
\label{fig:SDSSimage}
\end{figure}

\end{document}